\documentclass[aps,prc,twocolumn,showpacs,superscriptaddress,groupedaddress]{revtex4-1}
\usepackage{latexsym}
\usepackage{epsfig}
\usepackage{float}
\usepackage{lipsum}
\usepackage{graphicx,times}
\usepackage{amssymb,amsmath}
\usepackage{color}
\usepackage{dcolumn}
\usepackage{epstopdf}
\usepackage{pgfplots}
\usepackage[colorlinks=true, urlcolor=blue, linkcolor=blue, citecolor=blue]{hyperref}
\newcommand{\be}{\begin{equation}}
\newcommand{\ee}{\end{equation}}

\begin{document}
\title{Structure of $\Lambda$(1405) resonance and 
$\gamma{p}\rightarrow K^{+}+(\pi\Sigma)^{0}$ reaction}
\author{S. Marri}
\author{M. N. Nasrabadi}
\affiliation{Faculty of Physics, University of Isfahan, 81746-73441, Isfahan, Iran}
\author{S. Z. Kalantari}
\affiliation{Department of  Physics, Isfahan University of Technology, Isfahan 84156-83111, Iran}
\date{\today}
\begin{abstract}
Three-body calculations of $K\bar{K}N$ system with quantum numbers $I=1/2$, $J^{\pi}=(\frac{1}{2})^{+}$ 
were performed. Using separable potentials for two-body interactions, we calculated the $\pi\Sigma$ mass 
spectra for the $(\bar{K}N)_{I=0}+K^{+}\rightarrow(\pi\Sigma)^{0}K^{+}$ reaction on the basis of three-body 
Alt-Grassberger-Sandhas equations in the momentum representation. In this regard, different types of 
$\bar{K}N-\pi\Sigma$ potentials based on phenomenological and chiral SU(3) approach are used. The possibility 
to observe the trace of $\Lambda(1405)$ resonance in $(\pi\Sigma)^{0}$ mass spectra was studied. Using the 
$\chi^{2}$ fitting, it was shown that the mass of $\Lambda$(1405) resonance is about 1417 $\mathrm{MeV}/c^{2}$.
\end{abstract}
\pacs{13.75.Jz, 14.20.Pt, 21.85.+d, 25.80.Nv}
\maketitle
\section{Introduction}
\label{intro}
An important problem in the strangeness sector of the nuclear physics is the interaction 
of the antikaon and nucleon. The $I=0$ channel of the $\bar{K}N$ interaction is 
dominated by $\Lambda(1405)$ resonance, which is also a challenging issue in the 
strange nuclear physics. Because of the light mass of $\Lambda(1405)$ resonance, 
it is difficult to describe it as a 3-quark system in a constituent quark model. 
Therefore, the people usually take it as a quasi-bound state in $\bar{K}N$ system. 
Dalitz and Tuan predicted the existence of $\Lambda(1405)$ resonance using experimental 
data of the $\bar{K}N$ scattering length~\cite{dali1,dali2}. The experimental evidence 
of that was reported in 1961 in $K^{-}p\rightarrow\pi\pi\pi\Sigma$ reaction~\cite{alsto}. 
The $\pi^{+}\Sigma^{-}$ spectrum in $K^{-}d\rightarrow\pi^{+}\Sigma^{-}n$ reaction was 
studied by Braun {\it et al.,} and a resonance energy was found at 1420 MeV~\cite{bra}. 
The $\pi^{\pm}\Sigma^{\mp}$ mass spectra in $\pi^{-}p\rightarrow{K}^{+}\pi\Sigma$ reaction 
was studied in Ref.~\cite{thom} and a $\Lambda$ mass of 1405 $\mathrm{MeV}$ was deduced. 
During the recent years, many efforts have been made to study the structure and nature 
of $\Lambda$(1405) resonance~\cite{kais,oset1,olle,oset2,aka1,yam1,jido,dote1,hyodo,esm2,esm1,shev1,shev2,kh1}. 
The theoretical studies based on chiral SU(3) dynamics for $\bar{K}N$ interaction suggest 
a two-pole structure for $\Lambda$(1405)~\cite{kais,oset1,olle,oset2,jido,hyodo} while, 
the phenomenological models~\cite{aka1,yam1,dote1,shev1,shev2} suggest a one-pole structure.  

In recent years the theoretical and experimental investigations have been pursued by different 
groups to study the nature and structure of the $\Lambda$(1405) 
resonance~\cite{mor1,mor2,lep1,lep2,ahn,sieb,e31,fil,jid2,miya,jid3,jam1,shot}. In the last 
few years, a large number of data on $\gamma{p}\rightarrow{K}^{+}(\pi\Sigma)$ reaction were 
reported by CLAS collaboration at Jefferson Laboratory~\cite{mor1,mor2}. The $\pi\Sigma$ mass 
spectra in different particle channel was determined in a wide energy range and with good 
resolution. For the first time the quantum numbers of the $\Lambda(1405)$ resonance were 
determined based on experimental measurements~\cite{mor3}. The CLAS data together with the 
kaonic hydrogen data by can be used as a benchmark of our analyzing the $\bar{K}N$ interaction. 
Several theoretical studies have been performed to analyze the CLAD data based on chiral 
SU(3)~\cite{roca1,roca2,naka,mai,mis} and phenomenological approach~\cite{has}. The CLAS data 
were analyzed in Refs.~\cite{roca1,roca2,naka,mai,mis} using chiral dynamics and a double-pole 
structure was predicted for $\Lambda$ resonance, while, in Ref.~\cite{has} a $\Lambda$ mass of 
1405 $\mathrm{MeV}$ is deduced from the same CLAS data.

\begin{figure*}[htb]
\includegraphics[scale=0.65]{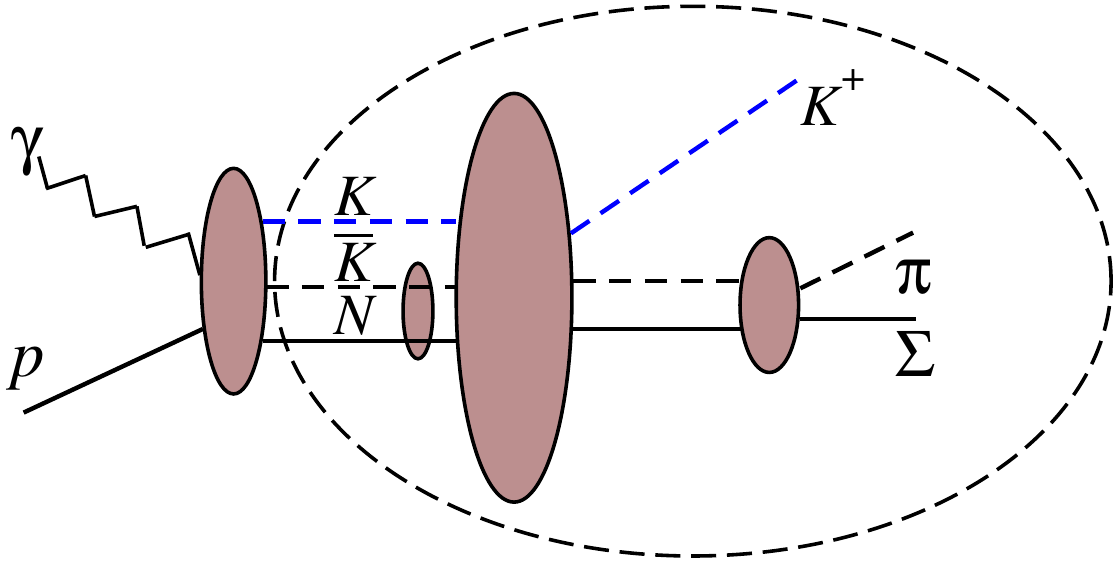}
\caption{Schematic depiction of the $\gamma{p}\rightarrow(\pi\Sigma)^{0}+K^{+}$ reaction.}
\label{fig.1}
\end{figure*}

The $\gamma{p}\rightarrow{K}^{+}(\pi\Sigma)^{0}$ reaction is depicted in Fig.~(\ref{fig.1}). 
In this reaction, after photon-proton interaction, we will have a $K\bar{K}N$ three-body 
system which can decay to the $K\pi\Sigma$ channel due to the Yukawa coupling of $K^{-}$ meson 
to the proton. In our approach, we decomposed the $\gamma{p}\rightarrow{K}^{+}(\pi\Sigma)$ 
reaction into two parts. The first part is $\gamma{p}\rightarrow{K}\bar{K}N$ reaction and the 
second part is $K\bar{K}N\rightarrow{K}^{+}(\pi\Sigma)$ reaction. In our calculations, we 
considered just the second part of the reaction. Thus, the processes in the dashed ellipse 
are included in the present work.

In this paper, the $\gamma{p}\rightarrow{K}^{+}(\pi\Sigma)^{0}$ reaction was studied by 
applying our approach based on the Alt-Grassberger-Sandhas (AGS) equations for $K\bar{K}N$ 
three-body system~\cite{alt}. In order to study the dependence of the results on two-body 
interactions, we used different types of phenomenological and chiral models of $\bar{K}N$ 
and $\bar{K}K$ interactions.

The paper is organized as follows. In Section~(\ref{formula}), a brief description of the Faddeev 
equations for $K\bar{K}N$ three-body system is given. The formula for calculating the $\pi\Sigma$ 
mass spectra and the two-body interactions, which are the inputs for the Faddeev equations and are 
given in Sec.~\ref{result}. The discussion of the results can also be found in Section~(\ref{result}). 
Finally, our conclusions will be summarized in Sec.~(\ref{conc}).
\section{Faddeev treatment of $K\bar{K}N$ three-body system}
\label{formula}
The three-body calculations of the $K\bar{K}N$ system is based on the AGS form 
of the Faddeev equations~\cite{alt}. We will use separable potentials for 
describing the two-body interactions with the form
\begin{equation}
V_{I}^{\alpha\beta}(k^{\alpha},k^{\beta};z)=g_{I}^{\alpha}(k^{\alpha})\, 
\lambda_{I}^{\alpha\beta}(z)\, g_{I}^{\beta}(k^{\beta}),
\label{eq.1}
\end{equation}
where $g_{I}^{\alpha}(k^{\alpha})$ is the form factor of the interacting two-body 
subsystem with relative momentum $k^{\alpha}$ and isospin $I$. The strength 
parameters of the interaction are shown by $\lambda_{I}^{\alpha\beta}$ and $z$ is 
a two-body energy. To include the lower lying channels in $\bar{K}N-\pi\Sigma$ and 
$\bar{K}K-\pi\pi-\pi\eta$ interactions, the potentials are labeled by $\alpha$ and 
$\beta$ indexes. The separable form of the two-body $T$-matrices are given by
\begin{equation}
T_{I}^{\alpha,\beta}(k^{\alpha},k^{\beta};z)=g_{I}^{\alpha}(k^{\alpha})\, 
\tau_{I}^{\alpha,\beta}(z)\, g_{I}^{\beta}(k^{\beta}),
\label{eq.2}
\end{equation}
where the operator $\tau_{I}^{\alpha,\beta}(z)$ is the usual two-body propagator. 

Using separable potential for two-body interactions, the three-body Faddeev 
equations~\cite{shev2} in the AGS form can be given by
\begin{equation}
\begin{split}
& \mathcal{K}_{i,j;I_{i},I_{j}}^{\alpha,\beta}(p_{i},p_{j};W)= (1-\delta_{ij})\, 
\mathcal{M}_{i,j;I_{i},I_{j}}^{\alpha}(p^{\alpha}_{i},p^{\alpha}_{j};W) \\
& +\sum_{\gamma}^{4}\sum_{k,I_{k}}\int {d}^{3}p_{k}\, \mathcal{M}_{i,k;I_i,I_k}^{\alpha}
(p^{\alpha}_{i},p^{\alpha}_{k};W)\, 
\tau_{k;I_k}^{\alpha,\gamma}(W-\frac{(p^{\alpha}_{k})^{2}}{2\nu^{\alpha}_{k}})\, \\
& \times\mathcal{K}_{k,j;I_{k},I_{j}}^{\gamma,\beta}(p^{\gamma}_{k},p^{\beta}_{j};W).
\label{eq.3}
\end{split}
\end{equation}

Here, the operators $\mathcal{K}_{i,j;I_{i},I_{j}}^{\alpha,\beta}$ are three-body transition 
amplitudes which describe the elastic and re-arrangement processes 
${[i+(jk)_{I_{i}}]}^{\alpha}\rightarrow {[j+(ki)_{I_{j}}]}^{\beta}$~\cite{shev2,kh2,kh3,kh4} 
and the operators $\mathcal{M}_{i,j;I_{i},I_{j}}^{\alpha}$ are Born terms which describe the 
effective potential realized by exchanged particle. In this equation, $W$ is the three-body 
energy and $W-\frac{(p^{\alpha}_{k})^{2}}{2\nu^{\alpha}_{k}}$ is the energy of the interacting 
pair $(ij)$ where, 
$\nu^{\alpha}_{k}=m^{\alpha}_{k}(m^{\alpha}_{i}+m^{\alpha}_{j})/(m^{\alpha}_{i}+m^{\alpha}_{j}+m^{\alpha}_{k})$, 
is the reduced mass, when particle $k$ is a spectator. Faddeev partition indexes $i,j,k=1,2,3$ 
denote simultaneously an interacting pair and a spectator particle. The operators in Faddeev 
Eq.~(\ref{eq.3}) are also labeled by particle indexes $\alpha,\beta$ and $\gamma=1,2,3,4$ to 
include the lower lying channels  
\begin{equation}
\begin{split}
& [\alpha=1]\, :K\bar{K}N, \hspace{1cm} [\alpha=2]\, :K\pi\Sigma,\\
& [\alpha=3]\, :\pi\pi{N}, \hspace{1.25cm} [\alpha=4]\, :\pi\eta{N}.
\end{split}
\label{eq.91}
\end{equation}
\section{RESULTS AND DISCUSSION}
\label{result}
Before we proceed to represent the discussion of the results, the two-body interactions 
should be defined. The two-body potentials are the basic ingredient of our three-body 
calculations for $K\bar{K}N$ system. Defining the subsystems of the three-body $K\bar{K}N$ 
system, the main two-body interactions are $\bar{K}N$, $K\bar{K}$ and $KN$. The $\bar{K}N$ 
subsystem is coupled with $\pi\Sigma$ and $K\bar{K}$ is coupled with $\pi\pi$ and $\pi\eta$ 
in $I=0$ and $I=1$ channels, respectively. Therefore, in the full Faddeev calculation of 
$K\bar{K}N$, one should also include the $K\pi\Sigma$, $\pi\pi{N}$ and $\pi\eta{N}$ channels. 
In the present work, firstly, the one channel AGS calculation was performed and the effect of 
the lower lying channels was taken into account effectively. Therefore, the two-body interactions 
in lower lying channels were neglected and decaying to $\pi\Sigma$, $\pi\pi$ and $\pi\eta$ was 
included by using the so-called exact optical potentials. In the present calculations, all 
potentials are in separable form having the form of Eq.~(\ref{eq.1}).

To describe the coupled channel $\bar{K}N-\pi\Sigma$ system, different models of interaction 
were used. We used four different phenomenological plus one chiral potential for $\bar{K}N$ 
interaction. The phenomenological potentials are energy-independent and have the one- and 
two-pole structures of the $\Lambda(1405)$ resonance. Therefore, in Eq.~(\ref{eq.1}), the strength 
parameters are energy-independent. The parameters of the phenomenological potentials, are given in 
Refs.~\cite{shev1,shev2}. In Ref.~\cite{shev1}, the potentials are adjusted to reproduce the 
results of the KEK experiment and in Ref.~\cite{shev2} the potentials are adjusted to reproduce 
the results of the SIDDHARTA experiment. From now on, we refer these phenomenological potentials as 
\textquotedblleft{KEK$^X$}\textquotedblright and \textquotedblleft{SIDD$^X$}\textquotedblright, 
respectively ($X=1,2$). For example, $\mathrm{SIDD}^{1}$ and $\mathrm{SIDD}^{2}$ are standing 
for a one- and a two-pole versions of the $\mathrm{SIDD}$ potential, respectively. The chiral 
potential is energy-dependent and reproduces a two-pole structure of $\Lambda(1405)$ resonance. 
The parameters of the chiral based potential are given in Ref.~\cite{ohnishi}. We refer to this 
potential as \textquotedblleft{Chiral}\textquotedblright. In Table~(\ref{ta.1}), the pole position 
of four different phenomenological (SIDD and KEK potentials) and one chiral based $\bar{K}N-\pi\Sigma$ 
interactions are presented. 
\begin{table*}[t]
\caption{The model dependence of the $\bar{K}N$ pole position(s) (in MeV). 
The pole position of four different phenomenological plus one chiral based 
$\bar{K}N-\pi\Sigma$ interactions are presented. In the first row 
$\mathrm{X}^{1}$ and $\mathrm{X}^{2}$ ($\mathrm{X}=\mathrm{SIDD},\mathrm{KEK}$) 
standing for a one- and a two-pole versions of the $\mathrm{SIDD}$ and 
$\mathrm{KEK}$ potentials, respectively.}
\centering
\begin{tabular}{cccccc}
\hline\hline\noalign{\smallskip}\noalign{\smallskip}
&\,  $\mathrm{SIDD}^{1}$ \, & \, $\mathrm{SIDD}^{2}$ \, & \, $\mathrm{KEK}^{1}$ \, & \, 
$\mathrm{KEK}^{2}$ \, & \, $\mathrm{Chiral}$  \\
\noalign{\smallskip}\noalign{\smallskip}\hline
\noalign{\smallskip}\noalign{\smallskip}
first pole \, & \, $1428.1-i46.6$ \, & \, $1418.1-i56.9$ \, & \, $1411.3-i35.8$ \, 
& \, $1410.8-i35.9$ \, & \, $1420.6-i20.3$ \\
second pole \, & \, $-$ \, & \, $1382.0-i104.2$ \, & \, $-$ \, & \, $1380.8-i104.8$ \, 
& \, $1343.0-i72.5$ \\
\noalign{\smallskip}\noalign{\smallskip}
\hline\hline
\end{tabular}
\label{ta.1} 
\end{table*}

For $KN$ interaction with isospin $I=0,1$, we used a one-channel real potential. The range parameters 
of the potentials were set to 3.9 $\mathrm{fm}^{-1}$ and the strength parameters are adjusted to 
reproduce the $KN$ scattering length. The experimental value of the scattering lengths for the $I=0$ and 
$I=1$ channels were set to $a^{I=0}_{KN}=-0.035\,\mathrm{fm}$ and $a^{I=1}_{KN}=-0.310\pm 0.003\,\mathrm{fm}$, 
respectively~\cite{dov,dum,jid4}. 

The $K\bar{K}$ interaction is attractive and together with its coupled channels dynamically 
generates the $f_{0}(980)$ and $a_{0}(980)$ resonances in $I=0$ and $I=1$ channels, respectively. 
We constructed our own potentials for the coupled-channel $K\bar{K}-\pi\pi$ and $K\bar{K}-\pi\eta$ 
interactions in the form of the Eq.~(\ref{eq.1}) with form-factors
\begin{equation}
g^{I}_{\alpha}(k_{\alpha})=\frac{1}{k^{2}_{\alpha}+(\Lambda^{I}_{\alpha})^{2}}.
\label{eq.92}
\end{equation}

To define the parameters $\lambda^{I}_{\alpha\beta}$ and $\Lambda^{I}_{\alpha}$, we used the mass and 
width of the $f_{0}$ and $a_{0}$ resonances, assuming that those are a quasi-bound state in the $K\bar{K}$ 
channel and a resonance in the $\pi\pi$ and $\pi\eta$ channels, respectively. For the pole-energy of 
$f_{0}$ and $a_{0}$ resonances, take the mass 980 MeV and the width 80 MeV. Therefore, the $f_{0}(980)$ 
($a_{0}(980)) $ pole is located on first Riemann sheet of the $K\bar{K}$ channel and in second Riemann 
sheet of the $\pi\pi$ ($\pi\eta$) channel. We also used the $K\bar{K}$ scattering length. For $K\bar{K}$ 
scattering length, we take the reported values in Ref.~\cite{bar} ($a_{K\bar{K}}^{I=0}=-0.14-i1.99\,\mathrm{fm}$ 
and $a_{K\bar{K}}^{I=1}=0.18-i0.61\,\mathrm{fm}$).
\subsection{The one-channel AGS calculation of the $K\bar{K}N-K\pi\Sigma-\pi\pi{N}-\pi\eta{N}$ system}
\label{dir1}
In the present subsection, the lower-lying channels of the $K\bar{K}N$ system has not been included 
directly and one-channel Faddeev AGS equations are solved for three-body $K\bar{K}N$ system and we 
approximated the full coupled-channel interaction by using the so-called exact optical $\bar{K}N(-\pi\Sigma)$ 
and $\bar{K}N(-\pi\pi-\pi\eta)$ potentials~\cite{shev1}. Therefore, the decaying to the $K\pi\Sigma$, 
$\pi\pi{N}$ and $\pi\eta{N}$ channels is taken into account through the imaginary part of the optical 
$\bar{K}N(-\pi\Sigma)$ and $\bar{K}N(-\pi\pi-\pi\eta)$ potentials. 

Since, the $K+(\bar{K}N)_{I=0}$ partition is the dominant state of the $K\bar{K}N$ three-body system, 
here, we supposed the $K+(\bar{K}N)_{I=0}$ partition as the initial state of the $K\bar{K}N$ system. 
Therefore, the Faddeev equations for $K+(\bar{K}N)_{I=0}$ reaction will have the form
\begin{widetext}
\begin{equation}
\begin{split}
& \mathcal{K}^{1,1}_{K,K;I_{{K}},0}\,\, =\,\, \hspace{1.75cm}\,\, +\,\, 
\mathcal{M}^{1}_{K,N;I_{K},I_{N}}\tau^{11}_{N;I_{N}}\mathcal{K}^{1,1}_{N,K;I_{N},0}
\hspace{0.05cm}\,\, +\,\, \mathcal{M}^{1}_{K,\bar{K};I_{K},I_{\bar{K}}}
\tau^{11}_{\bar{K};I_{\bar{K}}}\mathcal{K}^{1,1}_{\bar{K},K;I_{\bar{K}},0} \\
& \mathcal{K}^{1,1}_{N,K;I_{N},0}\,\, =\,\, \mathcal{M}^{1}_{N,K;I_{N},0}\,\, +\,\,\,\, 
\mathcal{M}^{1}_{N,K;I_{N},I_{K}}\tau^{11}_{K;I_{K}}
\mathcal{K}^{1,1}_{K,K;I_{K},0}\,\, +\,\, 
\mathcal{M}^{1}_{N,\bar{K};I_{N},I_{\bar{K}}}\tau^{11}_{\bar{K};I_{\bar{K}}}
\mathcal{K}^{1,1}_{\bar{K},K;I_{\bar{K}},0} \\
& \mathcal{K}^{1,1}_{\bar{K},K;I_{\bar{K}},0} \hspace{0.cm}\,\, =\,\,  
\mathcal{M}^{1}_{\bar{K},K;I_{\bar{K}},0}
\hspace{0.cm}\,\, +\,\, 
\mathcal{M}^{1}_{\bar{K},K;I_{\bar{K}},I_{K}}\tau^{11}_{K;I_{K}}\mathcal{K}^{1,1}_{K,K;I_{K},0}
\hspace{0.cm}\,\, +\,\, 
\mathcal{M}^{1}_{\bar{K},N;I_{\bar{K}},I_{N}}\tau^{11}_{N;I_{N}}\mathcal{K}^{1,1}_{N,K;I_{N},0}. \\
\end{split}
\label{eq.4}
\end{equation}
\end{widetext}

To study the possible signature of the $\Lambda(1405)$ resonance in the $\pi\Sigma$ mass spectra in the 
$K+(\bar{K}N)_{I=0}\rightarrow{K}\pi\Sigma$ reaction, first one should define scattering amplitude. As 
we do not include the lower lying channels directly into the calculations, the only Faddeev amplitude, 
which contribute in the scattering amplitude is $\mathcal{K}^{1,1}_{K,K;I_{K},0}(p_{K},P_{K};W)$ ($I_{K}=0,1$). 
Therefore, the scattering amplitude can be expressed as 
\begin{equation}
\begin{split}
& T_{(\pi\Sigma)+{K}\leftarrow(\bar{K}N)_{I=0}+K} (\vec{k}_{K},\vec{p}_{K},\vec{P}_{K};W) \\
& =\sum_{I_{K}}\, g_{2}^{I_{K}}(\vec{k}_{K})\, \tau^{21}_{K;I_{K}}
(W-\frac{p^{2}_{K}}{2\nu_{K}})\, \mathcal{K}^{1,1}_{K,K;I_{K},0}(p_{K},P_{K};W),
\end{split}
\label{eq.5}
\end{equation}
where, $\vec{k}_{i}$ is the relative momentum between the interacting pair ($jk$) and $\vec{P}_{K}$ 
is the initial momentum of the spectator $K$ in $K\bar{K}N$ center of mass. The quantity 
$\mathcal{K}^{1,1}_{i,j;I_{i},I_{j}}$ is the Faddeev amplitude, driven from equation~(\ref{eq.4}). 

Using Eq.~(\ref{eq.5}), we define the cross section of $K+(\bar{K}N)_{I=0}\rightarrow{K}+(\pi\Sigma)^{0}$ 
scattering as follows,
\begin{equation}
\begin{split}
& \frac{d\sigma}{dE_{K}}=\frac{\omega_{(\bar{K}N)} \omega_{K}}{W P_{K}}
\frac{m_{\pi}m_{\Sigma}m_{K}}{m_{\pi}+m_{\Sigma}+m_{K}} 
\int{d}\Omega_{p_{K}}d\Omega_{k_{K}}p_{K}k_{K} \\
& \times  \sum_{if}
|T_{(\pi\Sigma)^{0}+{K}\leftarrow(\bar{K}N)_{I=0}+K}(\vec{k}_{K},\vec{p}_{K},\vec{P}_{K};W)|^2,
\end{split}
\label{eq.6}
\end{equation}
where $E_{K}$ is the kaon energy in the center-of-mass frame and is defined by
\begin{equation}
E_{K}=m_{K}+\frac{p^{2}_{K}}{2\nu'_{K}}, \,\, 
\nu'_{K}=\frac{m_{K}(m_{\pi}+m_{\Sigma})}{m_{K}+m_{\pi}+m_{\Sigma}}
\label{eq.7}
\end{equation}
and the energies $\omega_{(\bar{K}N)}$ and $\omega_{K}$ are the kinetic energies of ${K}$ and 
$\bar{K}N$ in the initial state.
\begin{figure*}
\centering
\includegraphics[scale=0.28]{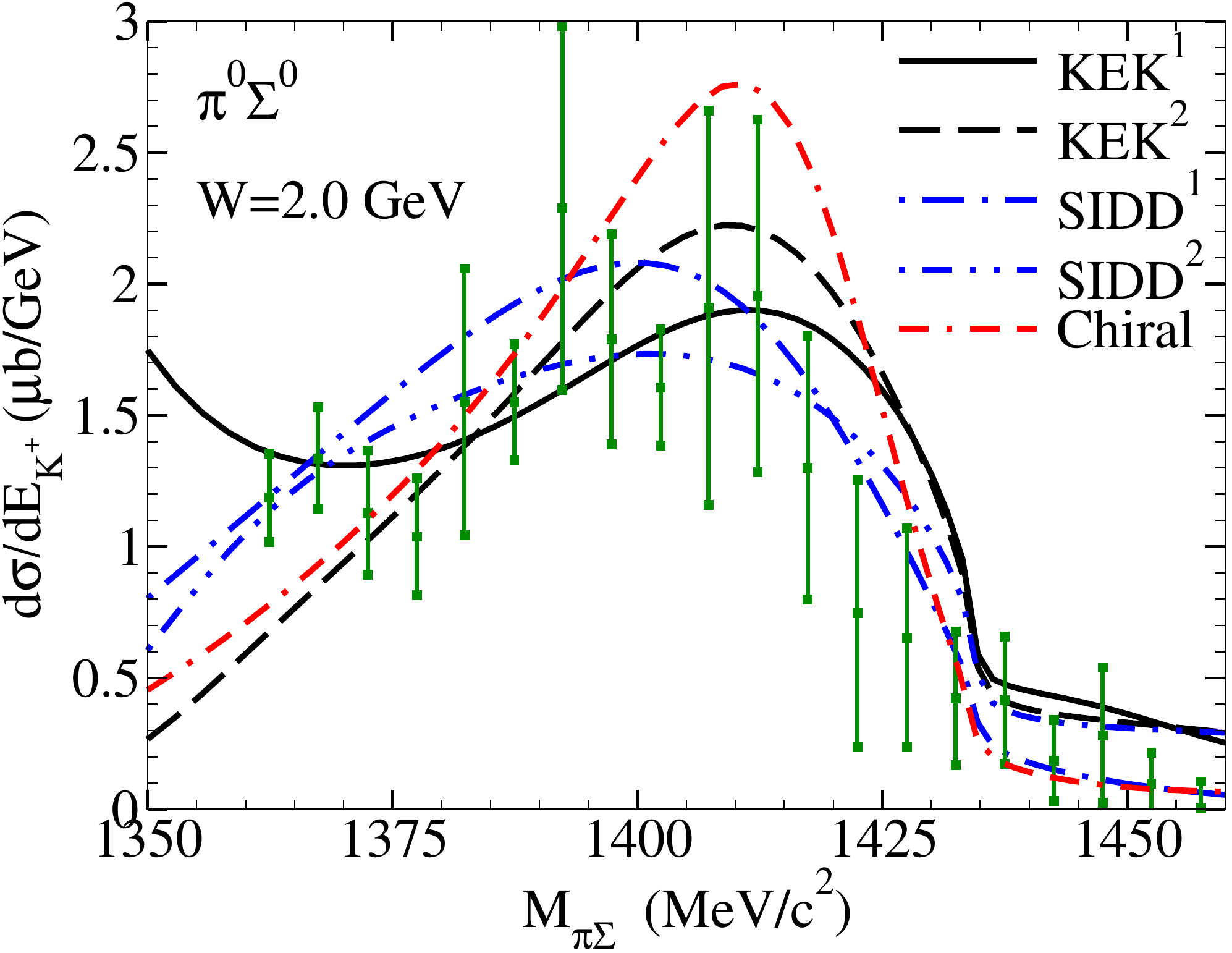}
\hspace{0.2cm}
\includegraphics[scale=0.28]{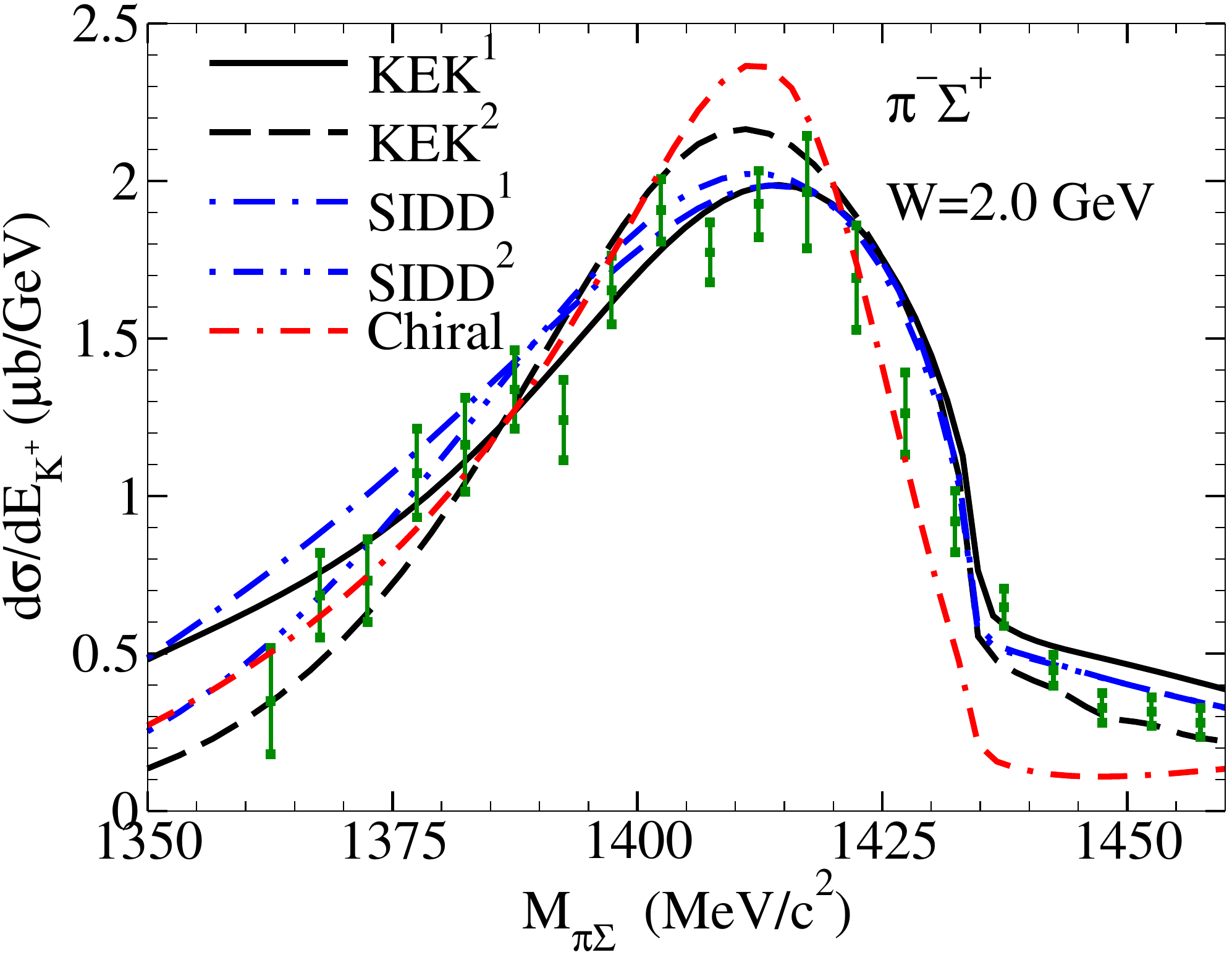}
\hspace{0.2cm}
\includegraphics[scale=0.28]{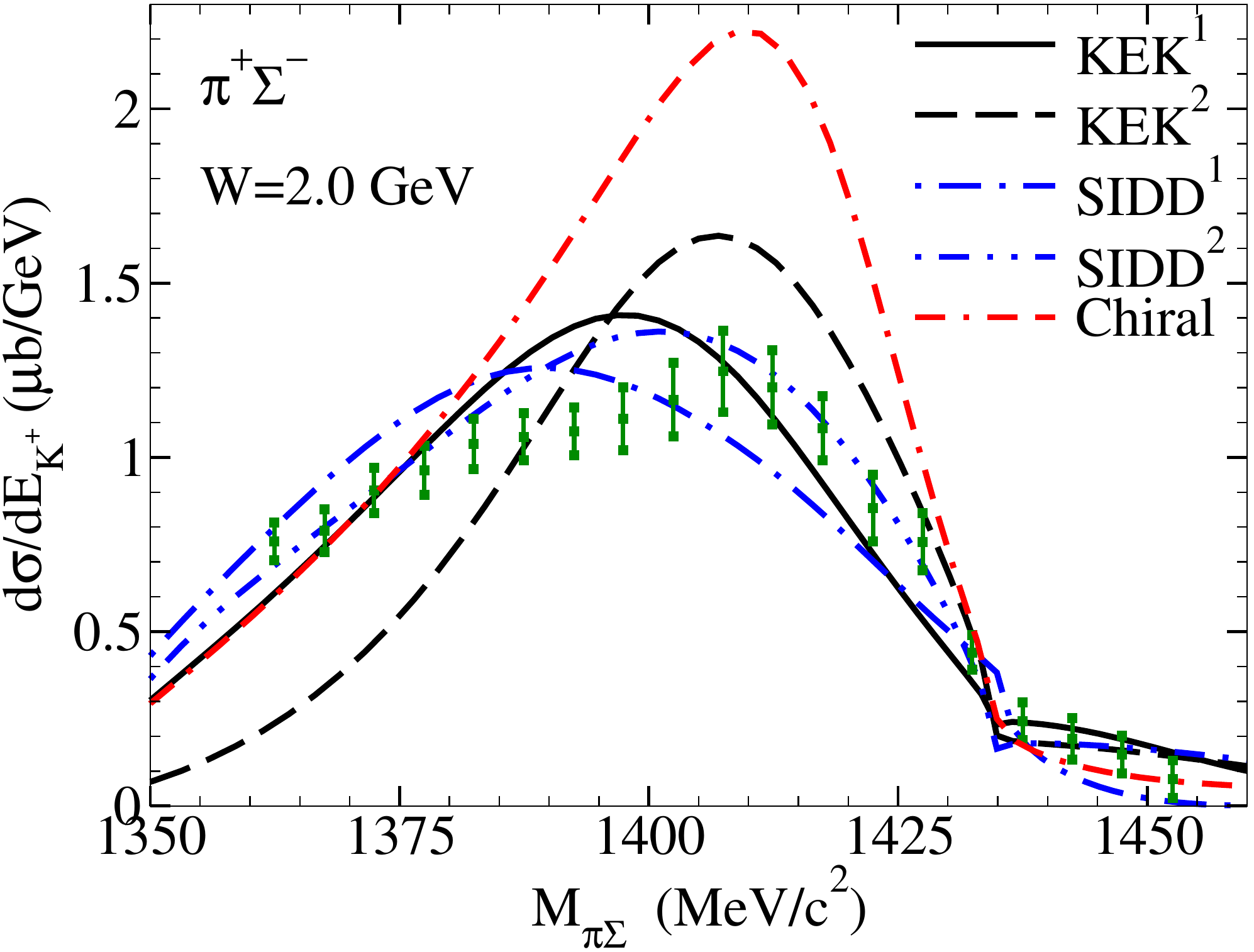} \\
\includegraphics[scale=0.28]{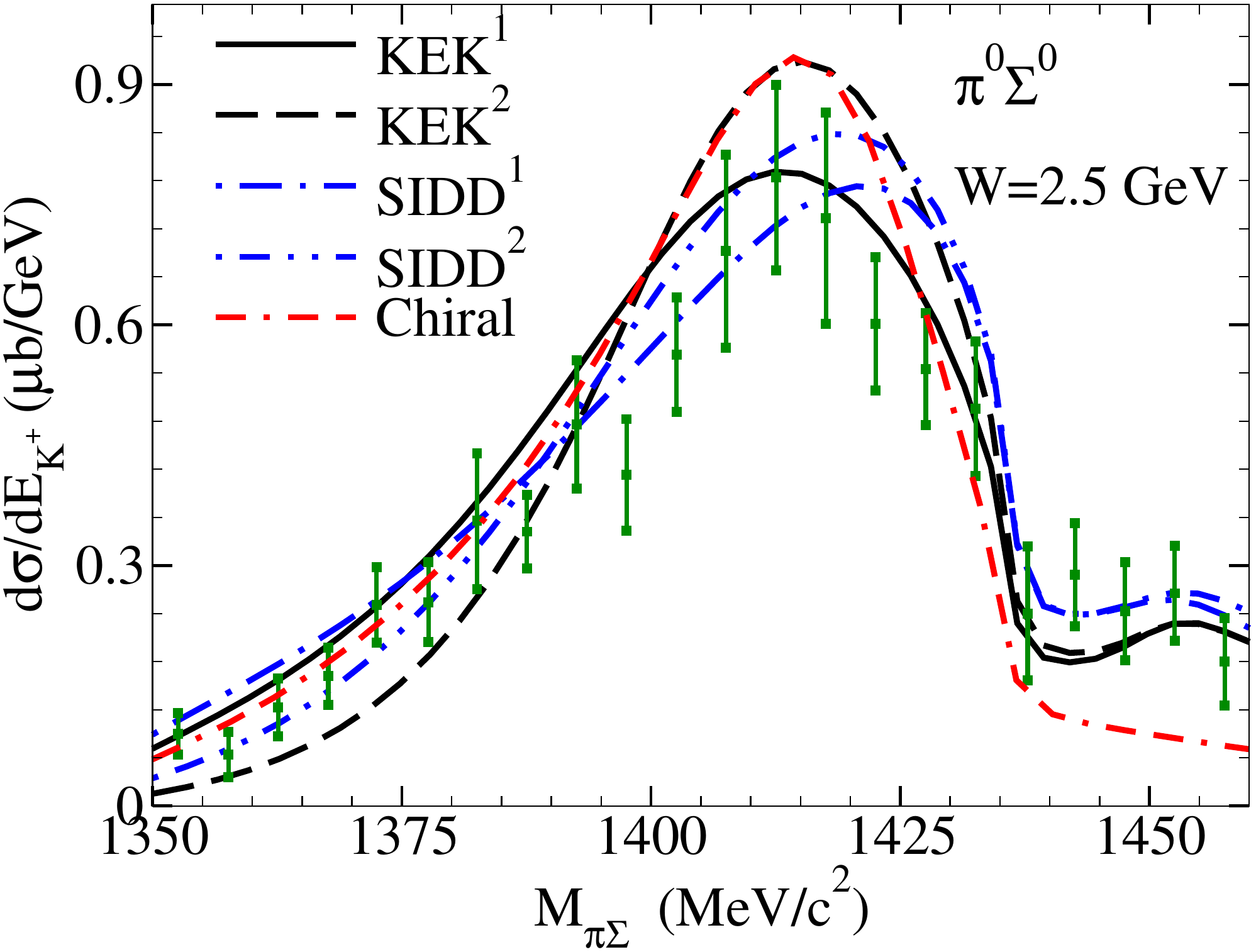}
\hspace{0.2cm}
\includegraphics[scale=0.28]{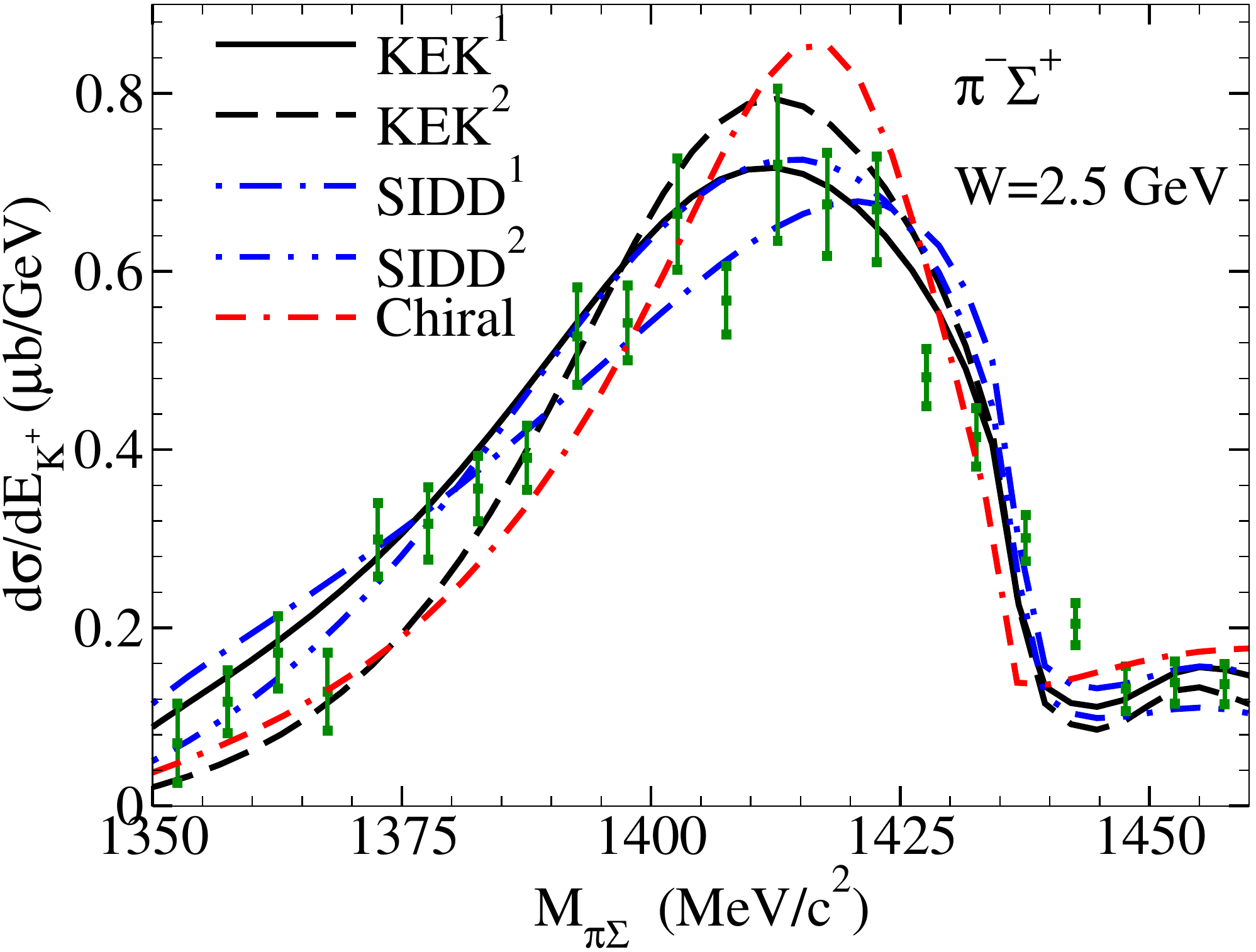}
\hspace{0.2cm}
\includegraphics[scale=0.28]{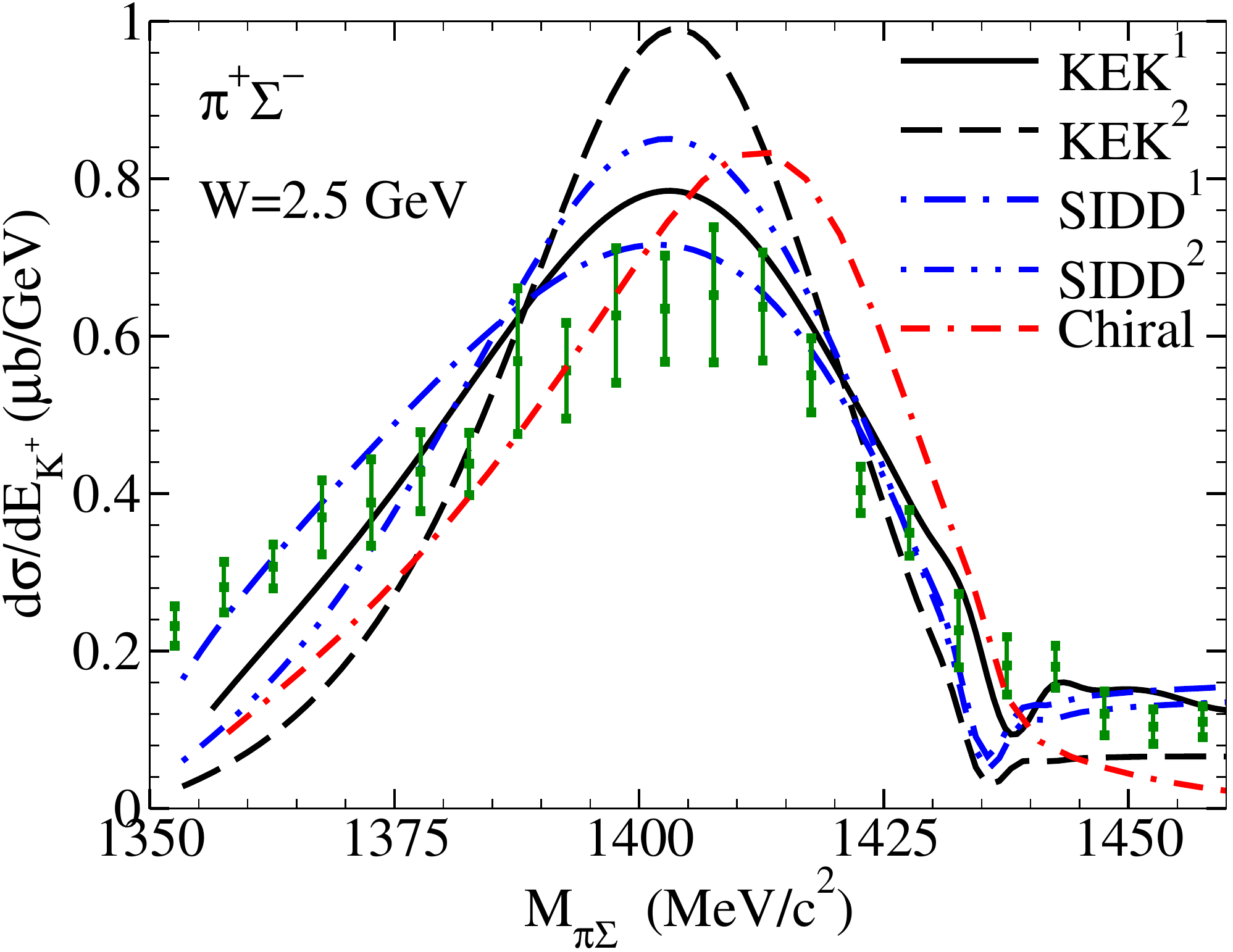} \\
\caption{(Color online) Comparison of the obtained results for $\pi\Sigma$ mass spectra in 
$K+(\bar{K}N)_{I=0}$reaction with the extracted experimental results~\cite{mor1,mor2} corresponding 
to $\gamma{p}\rightarrow(\pi\Sigma)^{0}+K^{+}$ reaction. Different types of $\bar{K}N-\pi\Sigma$ 
potentials were used. We used the one- and two-pole versions of the SIDD and KEK potentials~\cite{shev1,shev2} 
and also a chiral potential~\cite{ohnishi}. The initial energy of the system is around $E=2.0$ 
(upper row) and $E=2.5$ GeV (lower row). The black solid and black dashed lines show the mass spectra 
with one- and two-pole models of the KEK potential, respectively. The results for one- and two-pole 
models of the SIDD potential are depicted by the blue dash-dotted and blue dash-dot-dotted curves. 
Finally, the chiral results are represented by red dash-dash-dotted line.}
\label{fig.2}
\end{figure*}

Using Eq.~(\ref{eq.6}), we calculated the $\pi\Sigma$ mass spectra for $K+(\bar{K}N)_{I=0}$ reaction. 
The calculated $\pi\Sigma$ mass spectra for different $\bar{K}N$ model of interaction are depicted in 
Fig.~(\ref{fig.2}). In addition to the calculated mass spectra, the extracted data of CLAS experiment 
are also presented. The CLAS data are accessible for several energy bins, but in the present work, we 
calculated the $\pi\Sigma$ mass spectrum for two values of energy. In the upper row, the total energy 
of the three-body system in center of mass frame is $W=2\,\mathrm{GeV}$ and in the lower row the total 
energy of the system is $W= 2.5\,\mathrm{GeV}$. We used one-pole and two-pole versions of the KEK and SIDD 
potentials and also chiral based interaction to describe the $\bar{K}N$ interaction. The $\pi\Sigma$ mass 
spectra are calculated for all combinations of charges, i.e., $\pi^{+}\Sigma^{-}$, $\pi^{-}\Sigma^{+}$ and 
$\pi^{0}\Sigma^{0}$.

Since, the input energy of the AGS equations is real. Therefore, the moving singularities which are caused 
by the open channels, will appear in the three-body amplitudes. To remove the moon-shaped singularities, we 
followed the same procedure implemented in Refs.~\cite{poin1,poin2}. 

Comparing the experimental and theoretical results presented in Fig.~(\ref{fig.2}), we calculated the 
$\chi^{2}$ value for each model of $\bar{K}N$ interaction. 
\begin{equation}
\chi^{2}=\sum_{i=1}^{N}\frac{(N_{i}-\frac{d\sigma}{dE_{K}})^{2}}{(\sigma_{i})^{2}},
\label{eq.8}
\end{equation}
where $N$ denotes the number of data points which are included in the present fitting process. 
Here, $\frac{d\sigma}{dE_{K}}$ and $N_{i}$ are denoting the theoretical and experimental values for 
$K+(\bar{K}N)_{I=0}$ differential cross sections at energy $W$. The calculated values for $\chi^{2}$, 
are given in Table~(\ref{ta.2}) (the columns with (A) symbol, present the results by one-channel AGS 
calculations). In our calculations, we used 18 points of experimental results, starting from 
$M_{\pi\Sigma}=1352\,\mathrm{MeV}/c^{2}$ to $1437\,\mathrm{MeV}/c^{2}$.

\begin{figure*}
\centering
\includegraphics[width=8.5cm]{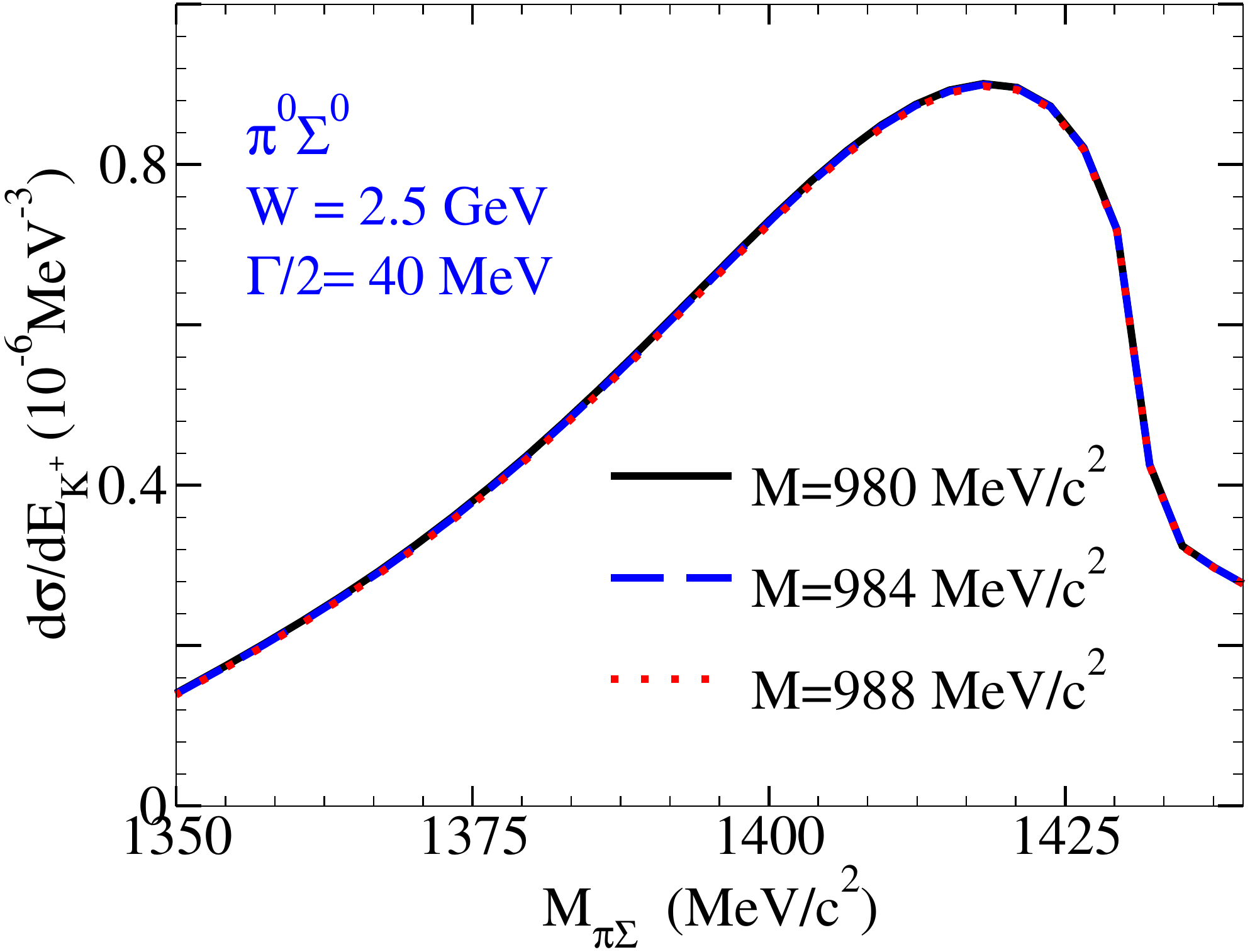} 
\hspace{0.5cm}
\includegraphics[width=8.5cm]{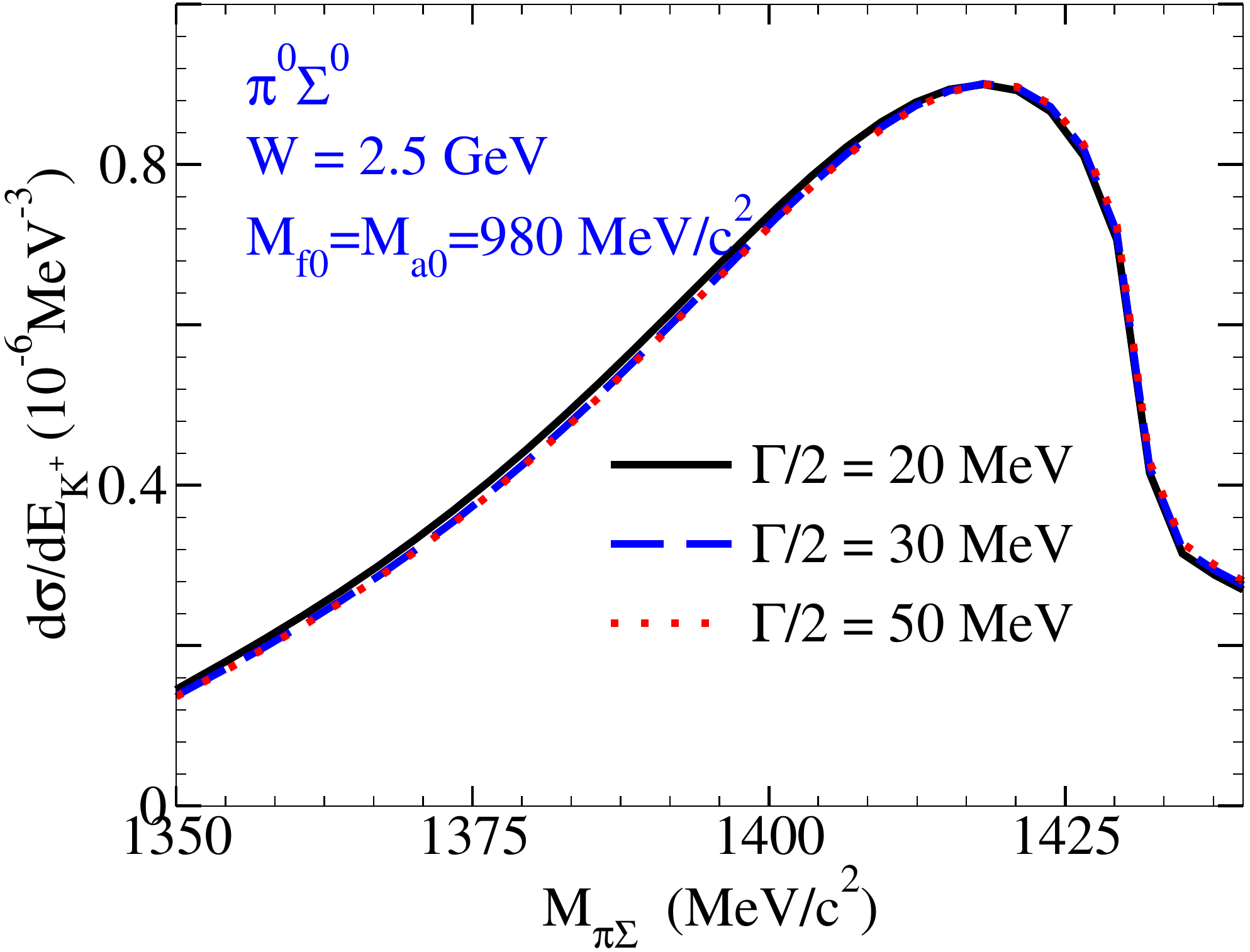}
\caption{(Color online) The dependence of the $\pi^{0}\Sigma^{0}$ mass spectra 
on the real (upper panel) and imaginary part (lower panel) of the $\bar{K}K$ 
pole position.}
\label{fig.44}
\end{figure*}

In addition to the $\bar{K}N$ interaction, the $K\bar{K}$ interaction is important in studying the 
dynamics of the $K\bar{K}N$ and may affect the experimental observable related to this system. To 
study the dependence of the mass spectra to the $K\bar{K}$ model of interaction, we also constructed 
several energy-independent potential by varying the mass and width of $f_{0}(980)$ and $a_{0}(980)$ 
resonances. In Fig.~(\ref{fig.44}), the variation of the $\pi^{0}\Sigma^{0}$ mass spectrum with respect 
to the real (left panel) and imaginary (right panel) part of the $\bar{K}K$ pole position at energy 
$W=2.5\mathrm{GeV}$ is shown. In the left panel, the width of the $f_{0}(980)$ and $a_{0}(980)$ resonances 
is fixed to be 80 MeV and their mass were varying from 980 to 988 MeV$/c^{2}$. In the right panel the 
mass of the resonances is fixed at 980 MeV$/c^{2}$ and the width was changed. As one can see, from the 
Fig.~(\ref{fig.44}), the $f_{0}(980)$ and $a_{0}(980)$ pole position variation can not change the 
$\pi^{0}\Sigma^{0}$ mass spectrum, effectively.

\subsection{The full coupled-channel AGS calculation of the $K\bar{K}N-K\pi\Sigma-\pi\pi{N}-\pi\eta{N}$ system}
\label{dir2}
In subsection~\ref{dir1}, we solved the one-channel AGS equations for $K\bar{K}N$ system and the 
decaying to the lower lying channels is included by using the so-called exact optical potential 
for $\bar{K}N$ and $\bar{K}K$ interactions. In one-channel Faddeev calculations the effect of the 
$\tau_{\pi\Sigma\rightarrow\pi\Sigma}$ amplitude was excluded. Based on chiral unitary approach, 
the first and second pole of $\Lambda$(1405) have clearly different coupling nature to the meson-baryon 
channels; the higher energy pole dominantly couples to the $\bar{K}N$ channel, while the lower energy 
pole strongly couples to the $\pi\Sigma$ channel. Due to the different coupling nature of these 
resonances, the shape of the $\Lambda$(1405) spectrum can be different depending on the initial and 
final channels. In the $\bar{K}N\rightarrow\pi\Sigma$ amplitude, the initial $\bar{K}N$ channel gets 
more contribution from the higher pole with a larger weight. Consequently, the spectrum shape has a 
peak around 1420 MeV coming from the higher pole~\cite{os1,ol1}. This is obviously different from 
the $\pi\Sigma\rightarrow\pi\Sigma$ spectrum which is largely affected by the lower pole. Therefore, 
the extracted mass spectra in subsection~\ref{dir1} can not reproduce exactly what we would see in 
an experiment.

\begin{table*}[htb]
\caption{The dependence of the $\chi^{2}$ parameter on the $\bar{K}N$ model of interaction in 
different particle channel. The obtained results in subsections~\ref{dir1},~\ref{dir2} and \ref{dir3} 
are labled by A, B and C symbols, respectively.}
\centering
\begin{tabular}{cccccccccccccccc}
\hline\hline\noalign{\smallskip}
    \,\,& & KEK$^{1}$ & \,\,& & KEK$^{2}$ & \,\,& & SIDD$^{1}$ & \,\,& & SIDD$^{2}$ & \,\,& & Chiral &  \\
\noalign{\smallskip}
\hline
\noalign{\smallskip}
$\underline{W=2.0\,\mathrm{GeV}}$\,\, & (A) & (B) & (C)\,\, & (A) 
& (B) & (C)\,\, & (A) & (B) & (C)\,\, & (A) & (B) & (C)\,\, & (A) & (B) & (C) \\
\noalign{\smallskip}
$\chi^{2} (\pi^{0}\Sigma^{0})$ \,\,
& $30.5$ & $14.4$ & $17.5$ \,\,
& $68.1$ & $28.9$ & $26.2$ \,\,
& $62.8$ & $11.2$ & $10.6$ \,\,
& $23.8$ & $16.3$ & $11.3$ \,\,
& $52.2$ & $24.8$ & $28.8$ \,\, \\
\noalign{\smallskip}
$\chi^{2} (\pi^{-}\Sigma^{+})$ \,\,
& $42.5$ & $32.7$ & $26.8$ \,\,
& $79.0$ & $79.8$ & $48.2$ \,\,
& $27.9$ & $14.4$ & $18.9$ \,\,
& $26.3$ & $21.5$ & $22.3$ \,\,
& $316$  & $97.7$ & $37.7$ \,\, \\
\noalign{\smallskip}
$\chi^{2} (\pi^{+}\Sigma^{-})$ \,\,
& $82.7$ & $99.8$ & $85.0$ \,\,
& $343$  & $362$  & $227 $ \,\, 
& $95.0$ & $42.5$ & $26.0$ \,\,
& $31.2$ & $92.6$ & $81.2$ \,\,
& $313$  & $312$  & $232 $ \,\,  \\
\noalign{\smallskip}
\hline
\noalign{\smallskip}
$\underline{W=2.5\,\mathrm{GeV}}$\,\, & (A) & (B) & (C)\,\, & (A) 
& (B) & (C)\,\, & (A) & (B) & (C)\,\, & (A) & (B) & (C)\,\, & (A) & (B) & (C) \\
\noalign{\smallskip}
$\chi^{2} (\pi^{0}\Sigma^{0})$ \,\,
& $27.7$  & $17.4$ & $11.2$ \,\,
& $62.6$  & $49.7$ & $24.4$ \,\,
& $20.9$ & $13.9$ & $13.2$ \,\,
& $30.9$ & $31.2$ & $23.3$ \,\,
& $79.5$ & $12.2$ & $12.5$ \,\, \\
\noalign{\smallskip}
$\chi^{2} (\pi^{-}\Sigma^{+})$\,\,
& $80.6$  & $91.5$ & $39.0$ \,\,
& $168$ & $83.4$ & $46.0$ \,\,
& $82.3$ & $37.7$ & $16.9$ \,\,
& $106$ & $31.5$ & $32.5$ \,\,
& $208$  & $22.5$ & $43.1$ \,\, \\
\noalign{\smallskip}
$\chi^{2} (\pi^{+}\Sigma^{-})$\,\,
& $100.7$ & $137$ & $66.8$ \,\,
& $448$   & $478$ & $299$ \,\,
& $58.7$  & $198$ & $39.7$ \,\,
& $217$   & $330$ & $144$ \,\,
& $270$   & $102$ & $67.5$ \,\, \\
\noalign{\smallskip}
\hline\hline
\end{tabular}
\label{ta.2}
\end{table*}

In addition to the above mentioned reaction, we need an interaction model for $\Sigma{K}$, $\pi{K}$, 
$\pi{N}$ and $\eta{N}$ subsystems. To describe the $\Sigma{K}$ interaction with isospin, we used a 
one-term complex potential for $I=1/2$ channel and a real potential for $I=3/2$ channel. To define 
the parameters of the $\Sigma{K}$ interaction, we used the $\Sigma{K}$ scattering length and the 
pole energy of $N^{*}(1535)$ resonance ($z_{pole}=1543-i46$ MeV)~\cite{pot1}. To describe the $\pi{N}$ 
interaction, we used the potential given in Ref.~\cite{ohnishi}. To define the $\pi{K}$ interaction 
in $I=1/2$ channel, we used a complex potential reproducing the pole energy of $K^{*}(892)$ resonance 
($z_{pole}=770-i250$ MeV)~\cite{pot2}. The parameters of the real $\pi{K}$ potential in $I=3/2$ isospin channels, 
we used the $\pi{K}$ phase shifts~\cite{pot2}.

To calculate the $\pi\Sigma$ mass spectra for $K+(\bar{K}N)_{I=0}$ reaction, we solved the Faddeev 
equations for coupled-channel $K\bar{K}N-K\pi\Sigma-\pi\pi-\pi\eta$ system which have the form
\begin{widetext}
\begin{equation}
\begin{split}
& \mathcal{K}^{1,1}_{K,K;I_{{K}},0}\,\, =\,\, 
\mathcal{M}^{1}_{K,\bar{K};I_{K},I_{\bar{K}}}
\tau^{11}_{\bar{K};I_{\bar{K}}}\mathcal{K}^{1,1}_{\bar{K},K;I_{\bar{K}},0}
+
\mathcal{M}^{1}_{K,N;I_{K},I_{N}}\tau^{11}_{N;I_{N}}\mathcal{K}^{1,1}_{N,K;I_{N},0} \\
& \hspace{1.7cm} +\mathcal{M}^{1}_{K,N;I_{K},I_{N}}\tau^{13}_{N;I_{N}} 
\mathcal{K}^{3,1}_{N,K;I_{N},0} 
+ \mathcal{M}^{1}_{K,N;I_{K},I_{N}}\tau^{14}_{N;I_{N}}\mathcal{K}^{4,1}_{N,K;I_{N},0}\\
& \mathcal{K}^{1,1}_{\bar{K},K;I_{\bar{K}},0} \hspace{0.cm}\,\, =\,\, 
\mathcal{M}^{1}_{\bar{K},K;I_{\bar{K}},0}
+
\mathcal{M}^{1}_{\bar{K},K;I_{\bar{K}},I_{K}}\tau^{11}_{K;I_{K}}\mathcal{K}^{1,1}_{K,K;I_{K},0}
+
\mathcal{M}^{1}_{\bar{K},K;I_{\bar{K}},I_{K}}\tau^{12}_{K;I_{K}}\mathcal{K}^{2,1}_{K,K;I_{K},0} \\
& \hspace{1.7cm}+\mathcal{M}^{1}_{\bar{K},N;I_{\bar{K}},I_{N}}\tau^{11}_{N;I_{N}} 
\mathcal{K}^{1,1}_{N,K;I_{N},0} 
+ \mathcal{M}^{1}_{\bar{K},N;I_{\bar{K}},I_{N}}\tau^{14}_{N;I_{N}}\mathcal{K}^{4,1}_{N,K;I_{N},0} \\
& \mathcal{K}^{1,1}_{N,K;I_{N},0}\,\, =\,\, \mathcal{M}^{1}_{N,K;I_{N},0}
+ 
\mathcal{M}^{1}_{N,K;I_{N},I_{K}} \tau^{11}_{K;I_{K}} \mathcal{K}^{1,1}_{K,K;I_{K},0} 
+ 
\mathcal{M}^{1}_{N,\bar{K};I_{N},I_{\bar{K}}} \tau^{11}_{\bar{K};I_{\bar{K}}} 
\mathcal{K}^{1,1}_{\bar{K},K;I_{\bar{K}},0} \\
& \hspace{1.7cm}+\mathcal{M}^{1}_{N,K;I_{N},I_{K}} \tau^{11}_{K;I_{K}} 
\mathcal{K}^{1,1}_{K,K;I_{K},0} \\
& \mathcal{K}^{2,1}_{K,K;I_{N},0} \,\, = \,\, 
\mathcal{M}^{2}_{K,\pi;I_{K},I_{\pi}} \tau^{22}_{\pi;I_{\pi}} \mathcal{K}^{2,1}_{\pi,K;I_{\pi},0} 
+ 
\mathcal{M}^{2}_{K,\Sigma;I_{K},I_{\Sigma}} \tau^{22}_{\Sigma;I_{\Sigma}} 
\mathcal{K}^{2,1}_{\Sigma,K;I_{\Sigma},0} \\
& \mathcal{K}^{2,1}_{\pi,K;I_{N},0} \,\,\, = \,\, 
\mathcal{M}^{2}_{\pi,K;I_{K},I_{K}} \tau^{22}_{K;I_{K}} \mathcal{K}^{2,1}_{K,K;I_{K},0} 
+ 
\mathcal{M}^{2}_{\pi,\Sigma;I_{K},I_{\Sigma}} \tau^{22}_{\Sigma;I_{\Sigma}} 
\mathcal{K}^{2,1}_{\Sigma,K;I_{\Sigma},0}
+
\mathcal{M}^{2}_{\pi,K;I_{K},I_{K}} \tau^{21}_{K;I_{K}} \mathcal{K}^{1,1}_{K,K;I_{K},0}  \\
& \mathcal{K}^{2,1}_{\Sigma,K;I_{N},0} \,\, = \,\, 
\mathcal{M}^{2}_{\Sigma,K;I_{K},I_{K}} \tau^{22}_{K;I_{K}} \mathcal{K}^{2,1}_{K,K;I_{K},0} 
+
\mathcal{M}^{2}_{\Sigma,\pi;I_{K},I_{K}} \tau^{22}_{K;I_{\pi}} \mathcal{K}^{2,1}_{\pi,K;I_{\pi},0} 
+
\mathcal{M}^{2}_{\Sigma,K;I_{K},I_{K}} \tau^{21}_{K;I_{K}} \mathcal{K}^{1,1}_{K,K;I_{K},0} \\
& \bar{\mathcal{K}}^{3,1}_{\pi,K;I_{\pi},0} \,\,\, = \,\, 
\mathcal{M}^{3}_{\pi_{1},\pi_{2};I_{\pi_{1}},I_{\pi_{2}}} \tau^{33}_{\pi_{2};I_{\pi_{2}}} 
\bar{\mathcal{K}}^{3,1}_{\pi,K;I_{\pi},0} 
+
2\mathcal{M}^{3}_{\pi_{1},N;I_{\pi_{2}},I_{N}} \tau^{33}_{N;I_{N}} 
\bar{\mathcal{K}}^{3,1}_{N,K;I_{N},0}
+
2\mathcal{M}^{3}_{\pi_{1},N;I_{\pi_{2}},I_{N}} \tau^{31}_{N;I_{N}} 
\bar{\mathcal{K}}^{1,1}_{N,K;I_{N},0} \\
& \mathcal{K}^{3,1}_{N,K;I_{N},0} \, = \,\, 
\mathcal{M}^{3}_{N,\pi_{1};I_{N},I_{\pi_{1}}} \tau^{33}_{\pi_{1};I_{\pi_{1}}} 
\bar{\mathcal{K}}^{3,1}_{\pi,K;I_{\pi},0} \\
& \mathcal{K}^{4,1}_{\pi,K;I_{\pi},0} \,\,\, = \,\, 
\mathcal{M}^{4}_{\pi,\eta;I_{\pi},I_{\eta}} \tau^{44}_{\eta;I_{\eta}} 
\mathcal{K}^{4,1}_{\eta,K;I_{\eta},0} 
+
\mathcal{M}^{4}_{\pi,N;I_{\pi},I_{N}} \tau^{44}_{N;I_{N}} \mathcal{K}^{4,1}_{N,K;I_{N},0} 
+
\mathcal{M}^{4}_{\pi,N;I_{\pi},I_{N}} \tau^{41}_{N;I_{N}} \mathcal{K}^{1,1}_{N,K;I_{N},0} \\
& \mathcal{K}^{4,1}_{\eta,K;I_{\eta},0} \,\,\,\, = \,\, 
\mathcal{M}^{4}_{\eta,\pi;I_{\eta},I_{\pi}} \tau^{44}_{\pi;I_{\pi}} 
\mathcal{K}^{4,1}_{\pi,K;I_{\pi},0} 
+
\mathcal{M}^{4}_{\eta,N;I_{\eta},I_{N}} \tau^{44}_{N;I_{N}} \mathcal{K}^{4,1}_{N,K;I_{N},0} 
+
\mathcal{M}^{4}_{\eta,N;I_{\eta},I_{N}} \tau^{41}_{N;I_{N}} \mathcal{K}^{1,1}_{N,K;I_{N},0} \\
& \mathcal{K}^{4,1}_{N,K;I_{N},0} \, = \,\, 
\mathcal{M}^{4}_{N,\pi;I_{N},I_{\pi}} \tau^{44}_{\pi;I_{\pi}} \mathcal{K}^{4,1}_{\pi,K;I_{\pi},0} 
+
\mathcal{M}^{4}_{N,\eta;I_{N},I_{\eta}} \tau^{44}_{\eta;I_{\eta}} \mathcal{K}^{4,1}_{\eta,K;I_{\eta},0} 
\end{split}
\label{eq.9}
\end{equation}
\end{widetext}

Since, there are two identical pion in $\pi\pi{N}$ channel, the operators in this channel should be 
symmetrized. Therefore, the operators $\mathcal{K}^{3,1}_{\pi_{1},K;I_{\pi_{1}},0}$ and 
$\mathcal{K}^{3,1}_{\pi_{2},K;I_{\pi_{2}},0}$ are replaced by $\bar{\mathcal{K}}^{3,1}_{\pi,K;I_{\pi},0}$ 
which is given by
\begin{equation}
\bar{\mathcal{K}}^{3,1}_{\pi,K;I_{\pi},0}=\mathcal{K}^{3,1}_{\pi_{1},K;I_{\pi_{1}},0}
+\mathcal{K}^{3,1}_{\pi_{2},K;I_{\pi_{2}},0}.
\label{eq.10}
\end{equation}

The scattering amplitude for $K+(\bar{K}N)_{I=0}\rightarrow K+(\pi\Sigma)$ reaction in 
terms of the Faddeev transition amplitudes can be given by
\begin{equation}
\begin{split}
& T_{(\pi\Sigma){K}\leftarrow(\bar{K}N)_{I=0}+K} (\vec{k}_K,\vec{p}_K,P_{K};W) \\
& =\sum_{I_{K}}g^{2}_{K;I_{K}}(\vec{k}_K )\tau^{21}_{K;I_{K}}(W-E_K(\vec{p}_K))
\mathcal{K}^{1,1}_{K,K;I_{K},0}(p_{K},P_{K};W) \\
& +\sum_{I_{K}}g^{2}_{K;I_{K}}(\vec{k}_K )\tau^{22}_{K;I_{K}}(W-E_{K}(\vec{p}_{K})) 
\mathcal{K}^{2,1}_{K,K;I_{K},0}(p_{K},P_{K};W) \\
& +\sum_{I_{\pi}}\sum_{I_{K}}\langle[\pi\otimes\Sigma]_{I_{K}}\otimes{K}
\mid\pi\otimes[\Sigma\otimes{K}]_{I_{\pi}}
\rangle{g}^{2}_{\pi;I_{\pi}}(\vec{k}_{\pi}) \\
& \hspace{1.5cm}\times \tau^{22}_{\pi;I_{\pi}}(W-E_{\pi}(\vec{p}_{\pi})) 
\mathcal{K}^{21}_{\pi,K;I_{\pi},0}(p_{\pi},P_{K};W) \\
& +\sum_{I_{\Sigma}}\sum_{I_{K}}\langle[\pi\otimes\Sigma]_{I_{K}}\otimes{K}
\mid\Sigma\otimes[\pi\otimes{K}]_{I_{\Sigma}}
\rangle{g}^{2}_{\Sigma;I_{\Sigma}}(\vec{k}_{\Sigma}) \\
& \hspace{1.5cm}\times \tau^{22}_{\Sigma;I_{\Sigma}}(W-E_{\Sigma}(\vec{p}_{\Sigma})) 
\mathcal{K}^{21}_{\Sigma,K;I_{\Sigma},0}(p_{\Sigma},P_{K};W).
\end{split}
\label{eq.11}
\end{equation}

\begin{figure*}
\centering
\includegraphics[scale=0.28]{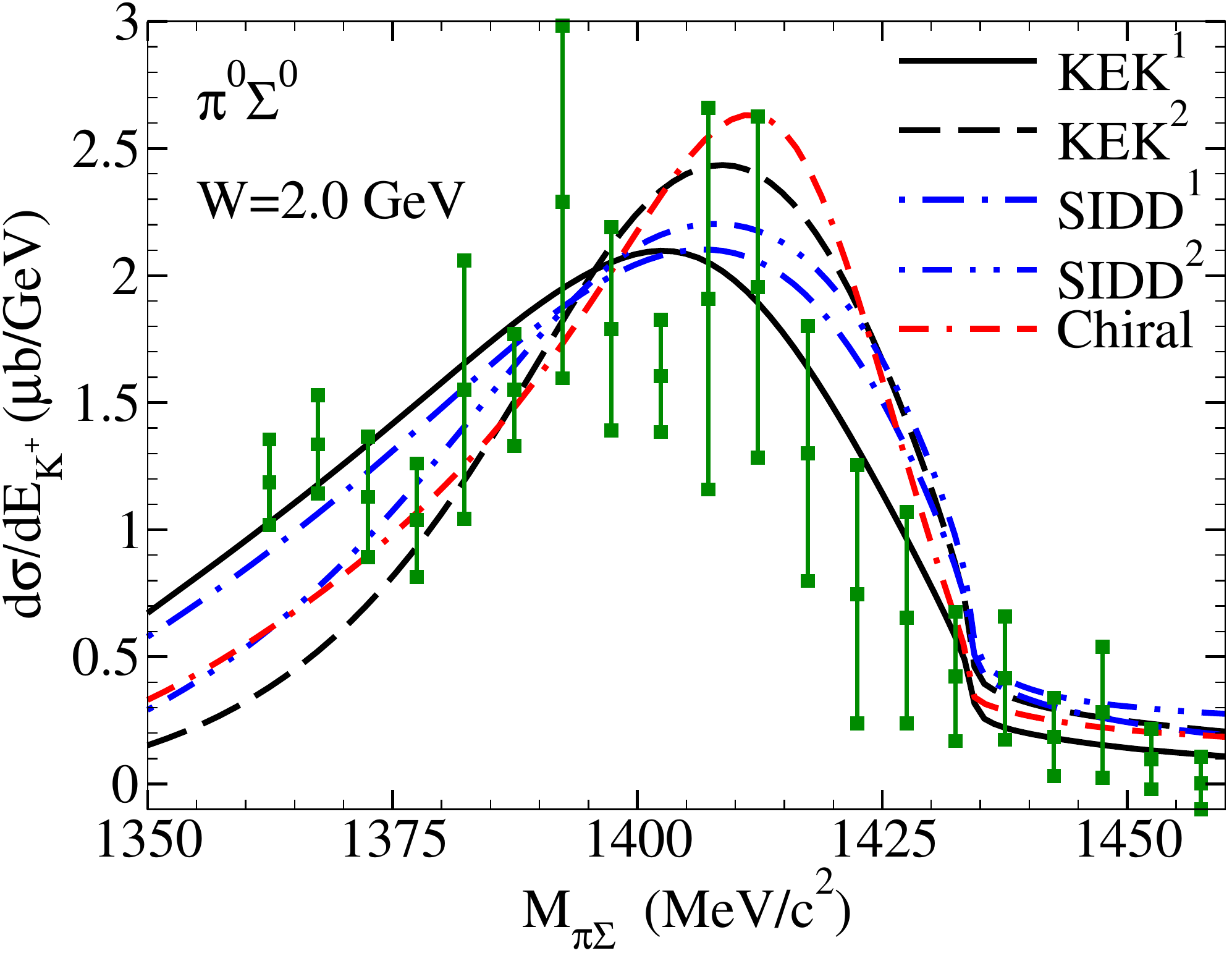}
\hspace{0.2cm}
\includegraphics[scale=0.28]{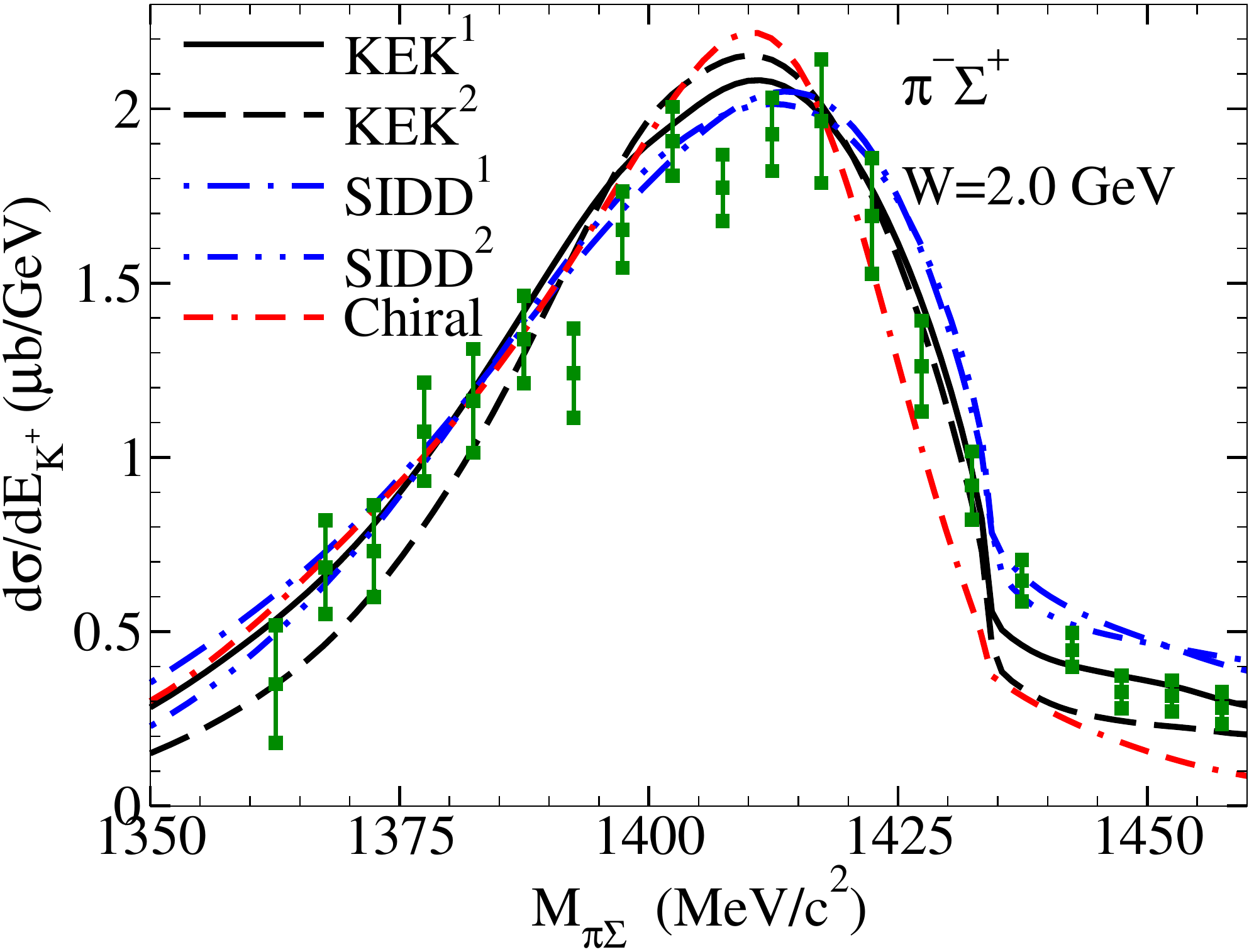}
\hspace{0.2cm}
\includegraphics[scale=0.28]{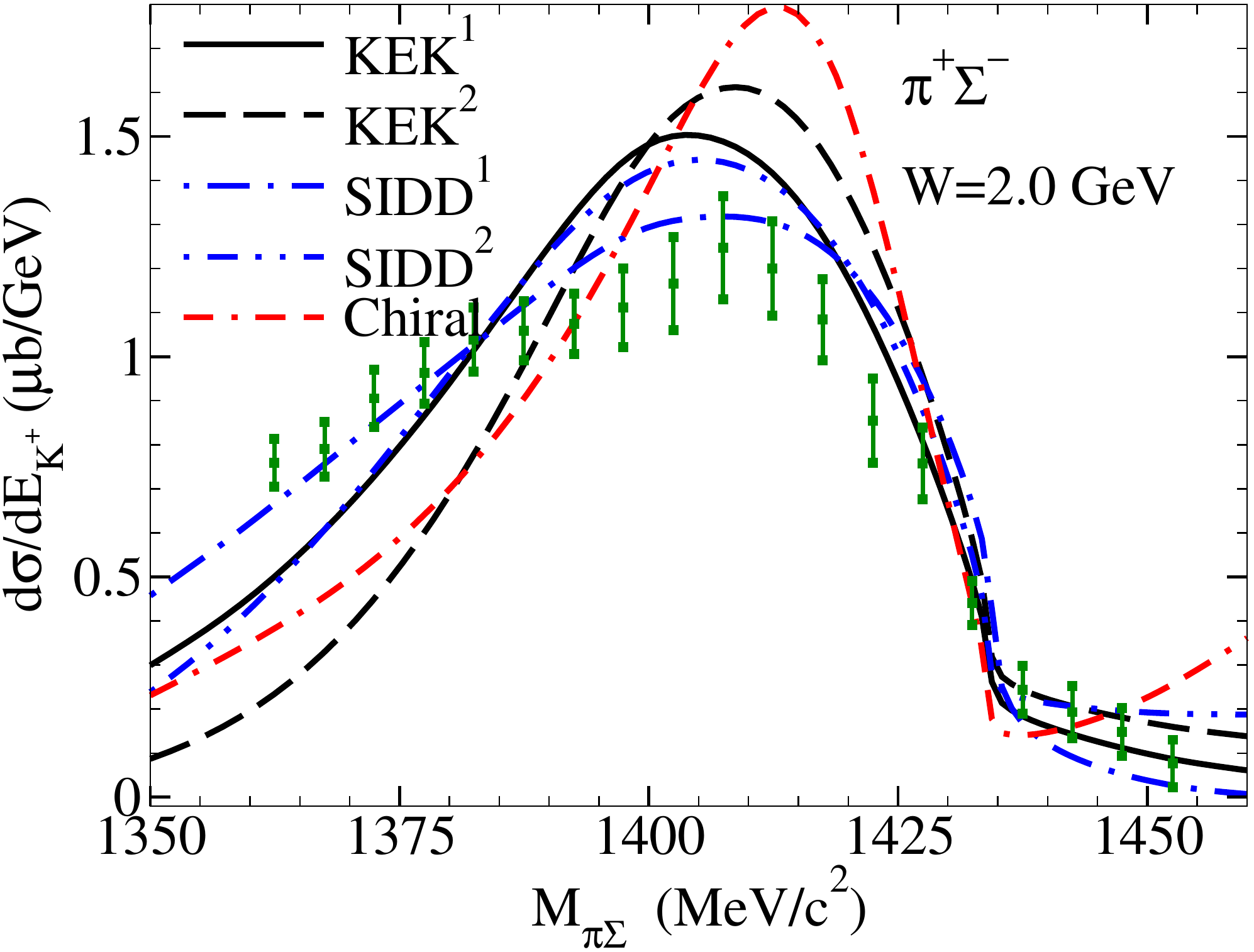} \\
\includegraphics[scale=0.28]{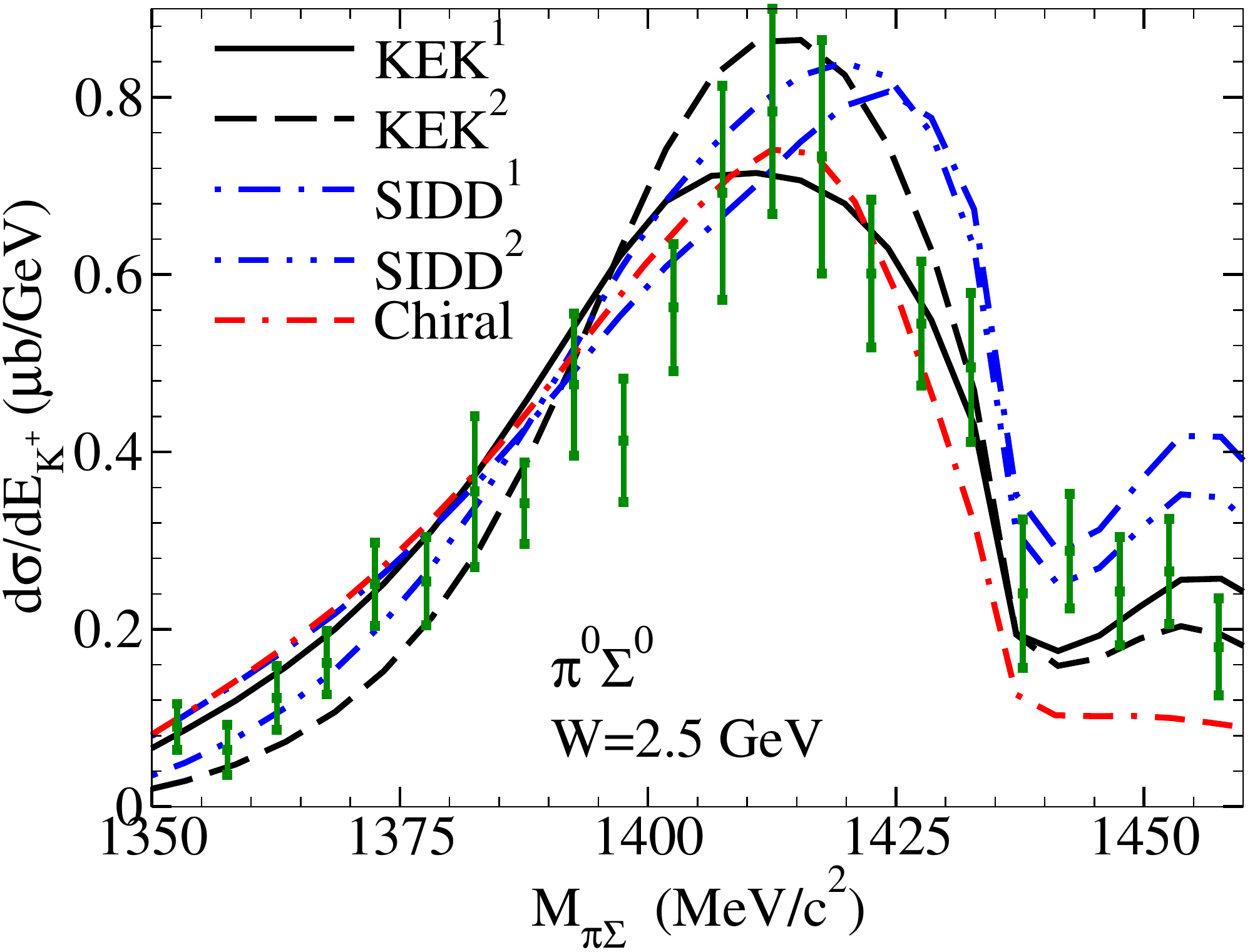}
\hspace{0.2cm}
\includegraphics[scale=0.28]{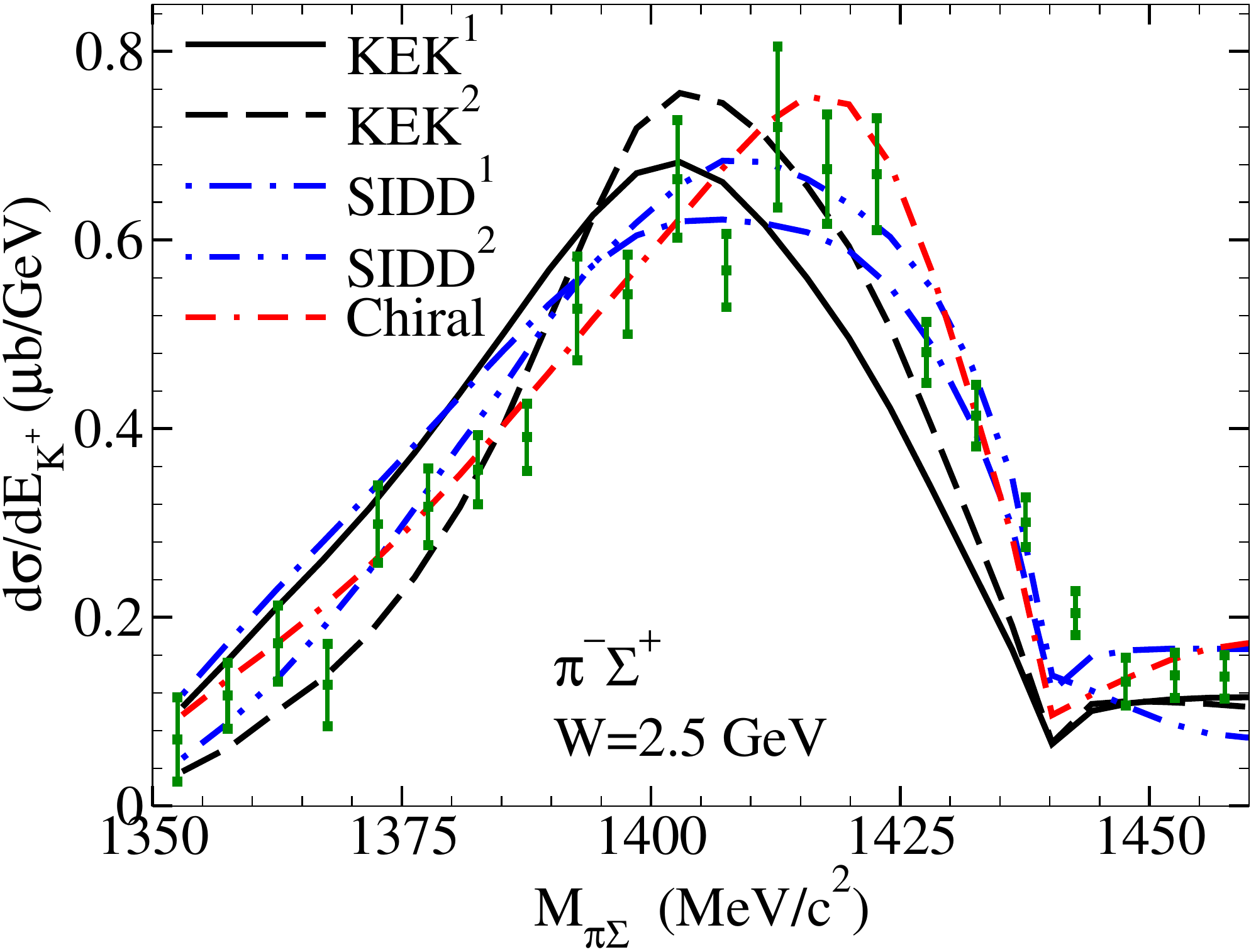}
\hspace{0.2cm}
\includegraphics[scale=0.28]{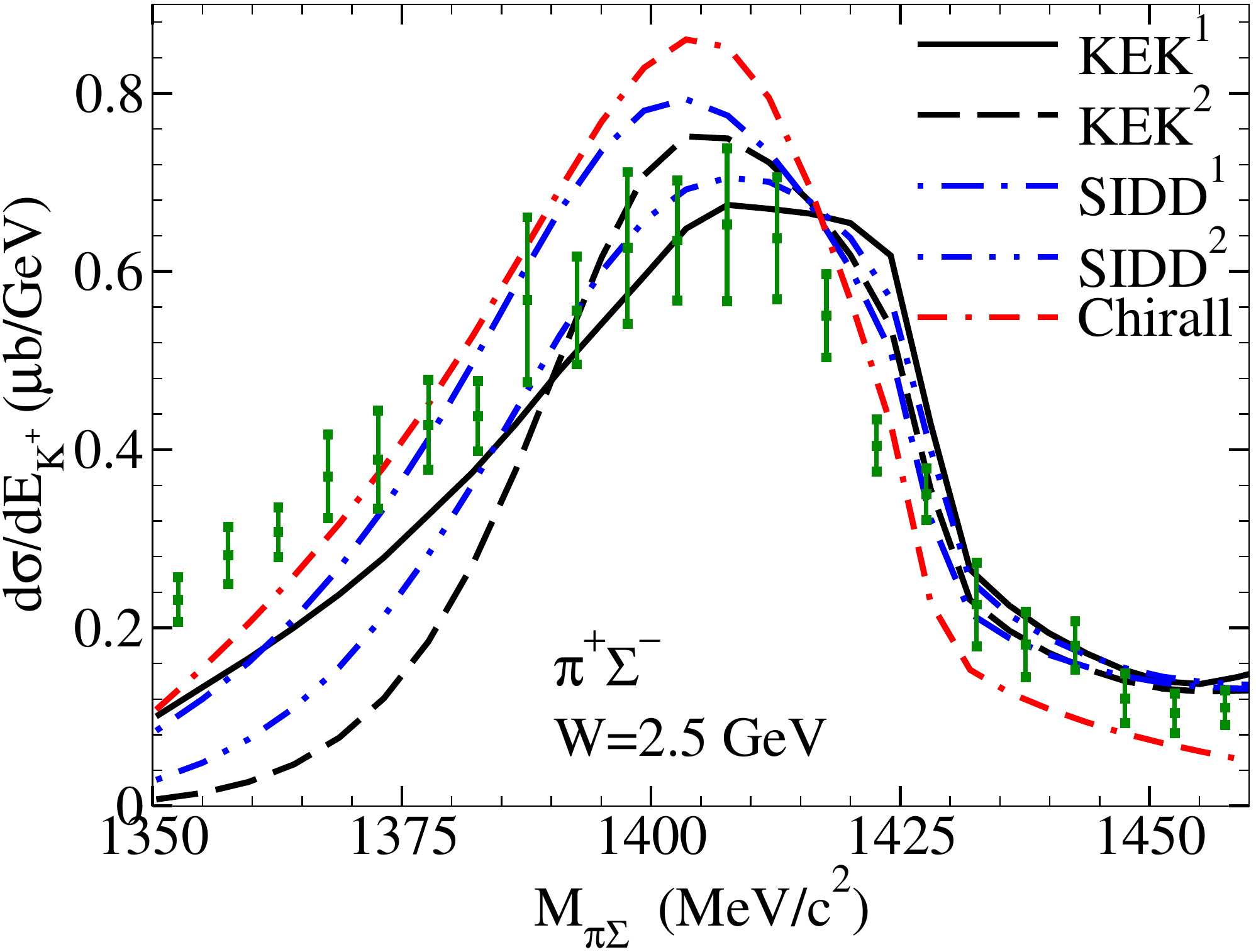} \\
\caption{(Color online) The $\pi\Sigma$ mass spectra for $\gamma{p}\rightarrow(\pi\Sigma)^{0}+K^{+}$ 
reaction. The descriptions are same as in Fig.~\ref{fig.2}, but in the present calculations, 
the full coupled-channel Faddeev AGS equations for $K\bar{K}N-K\pi\Sigma-\pi\pi{N}-\pi\eta{N}$ 
system are solved.}
\label{fig.3}
\end{figure*}

As one see from Eq.~\ref{eq.11}, in coupled-channel calculations plus the $\mathcal{K}^{1,1}_{K,K;I_{K},0}$ 
amplitude, the effect of the $\mathcal{K}^{2,1}_{K,K;I_{K},0}$, $\mathcal{K}^{21}_{\pi,K;I_{\pi},0}$ and 
$\mathcal{K}^{21}_{\Sigma,K;I_{\Sigma},0}$ are also included which accordingly, produces a more precise mass 
spectrum for $\pi\Sigma$. Inserting the new scattering amplitude (Eq.~\ref{eq.11}) in Eq.~\ref{eq.6}, we can 
calculate the $\pi\Sigma$ mass spectrum for $K+(\bar{K}N)_{I=0}\rightarrow K+(\pi\Sigma)^{0}$ reaction. In 
Fig.~\ref{fig.3}, we calculated the $\pi\Sigma$ mass spectrum using different potential models for 
$\bar{K}N-\pi\Sigma$ interaction. We also calculated the $\chi^{2}$ values for each model of interaction, 
which are presented in Table~\ref{ta.2} (the columns shown by (B) symbol). By comparison of results using 
one-channel and coupled-channel Faddeev equations, it may be possible to study the effect of 
$\tau_{\pi\Sigma\rightarrow\pi\Sigma}$ amplitude on $\pi\Sigma$ invariant mass. As can be seen in 
Table.~\ref{ta.2}, the inclusion of $\tau_{\pi\Sigma\rightarrow\pi\Sigma}$ amplitude can reduce the $\chi^{2}$ 
values especially, in $\pi^{0}\Sigma^{0}$ channel.

\begin{figure*}
\centering
\includegraphics[scale=0.28]{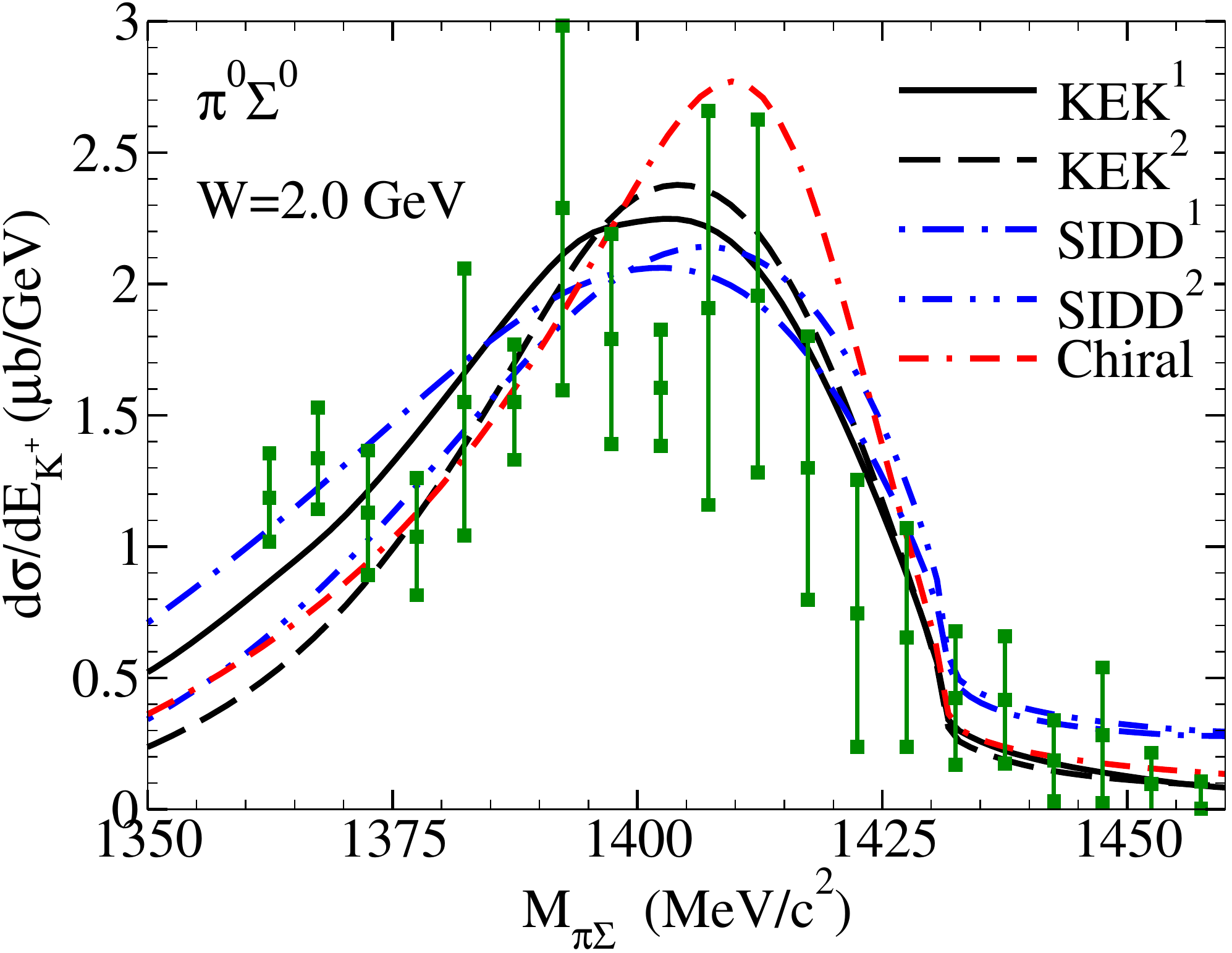}
\hspace{0.2cm}
\includegraphics[scale=0.28]{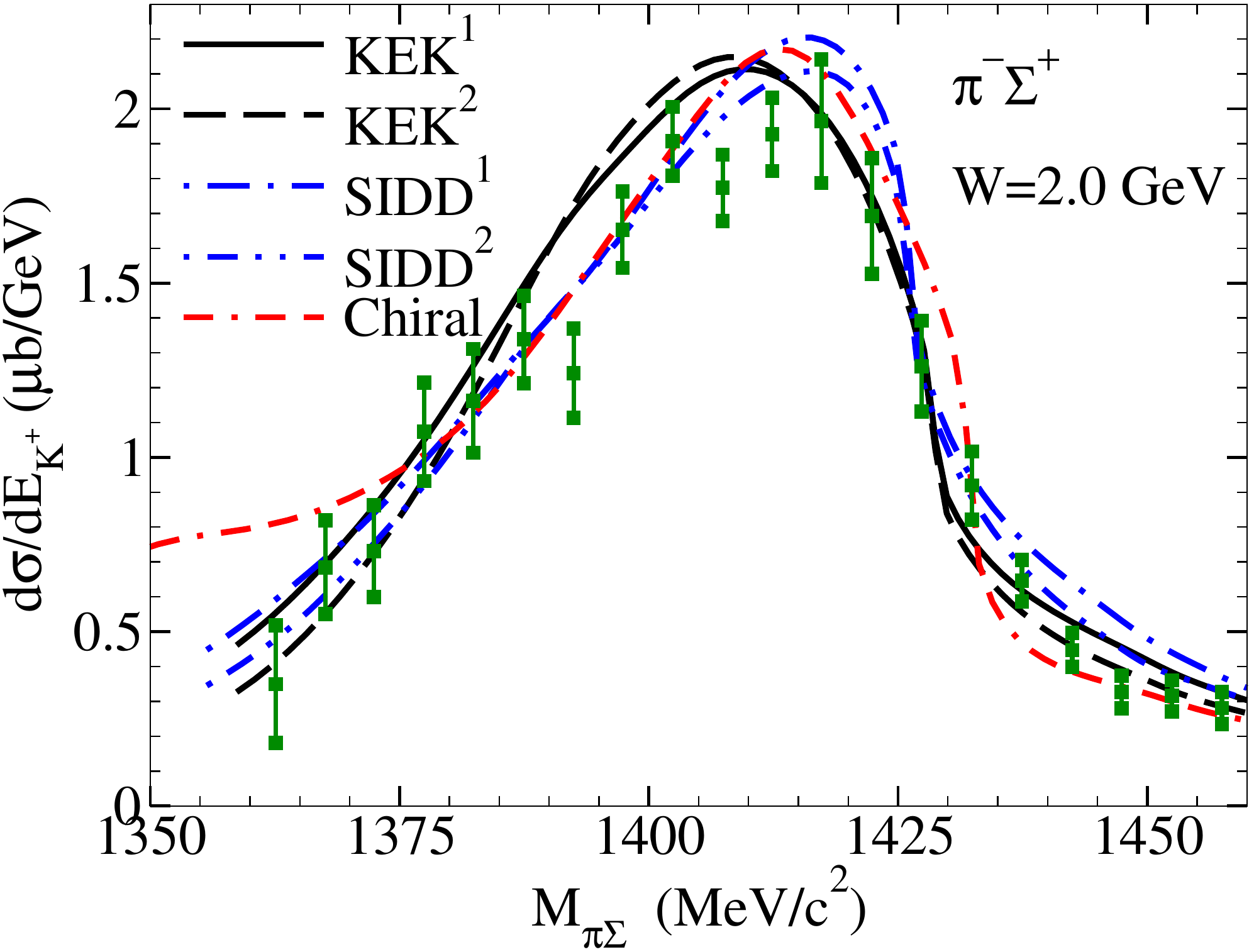}
\hspace{0.2cm}
\includegraphics[scale=0.28]{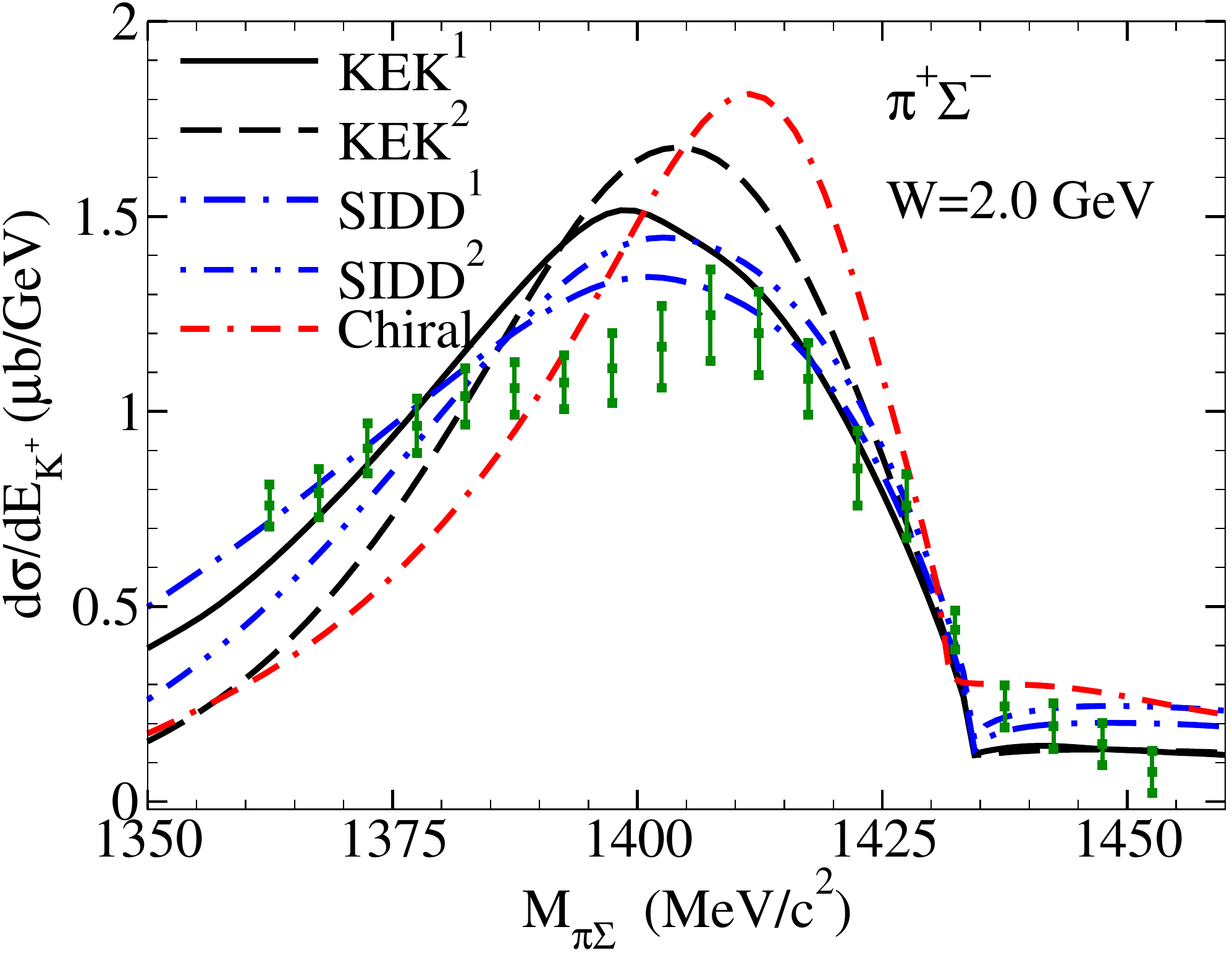} \\
\includegraphics[scale=0.28]{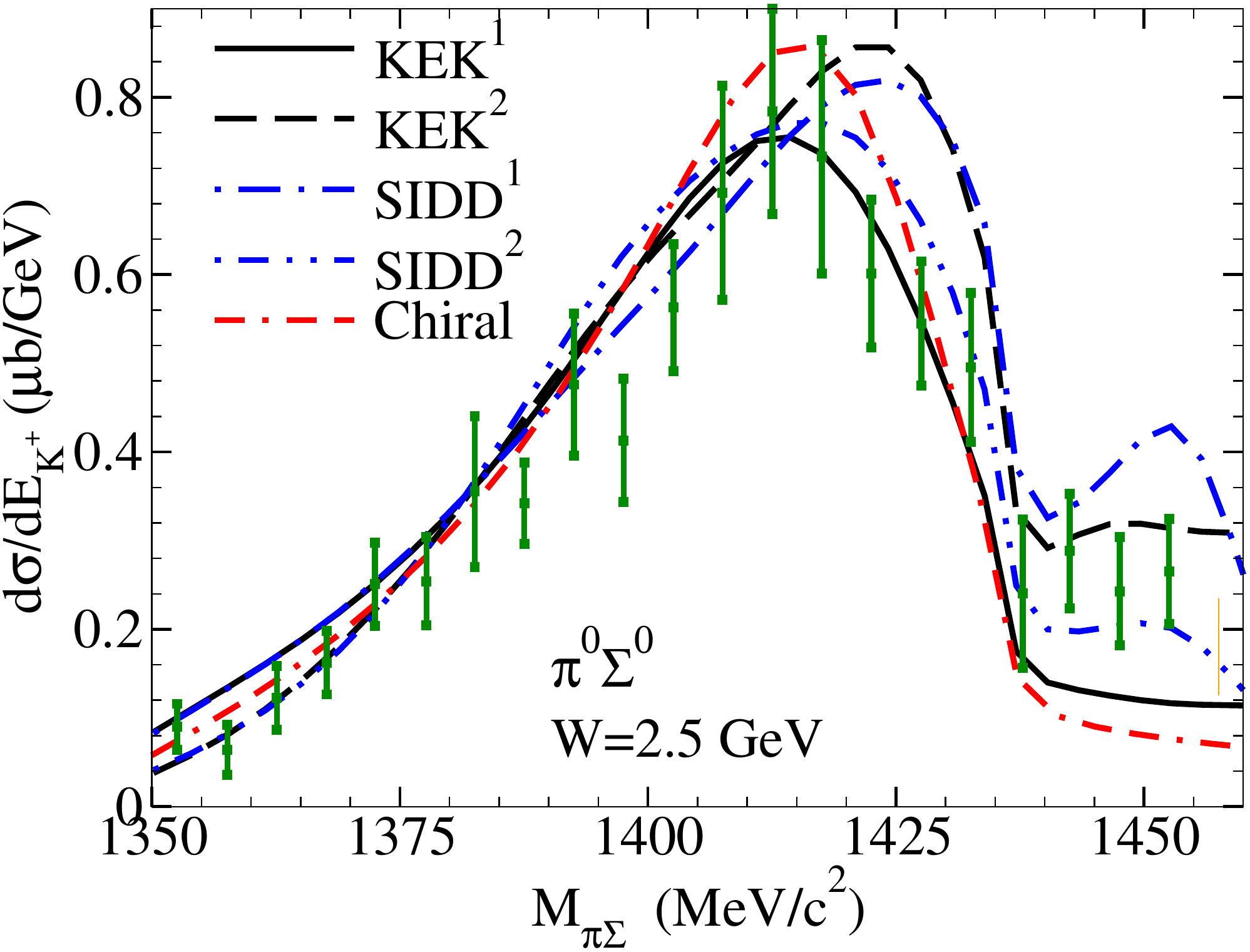}
\hspace{0.2cm}
\includegraphics[scale=0.28]{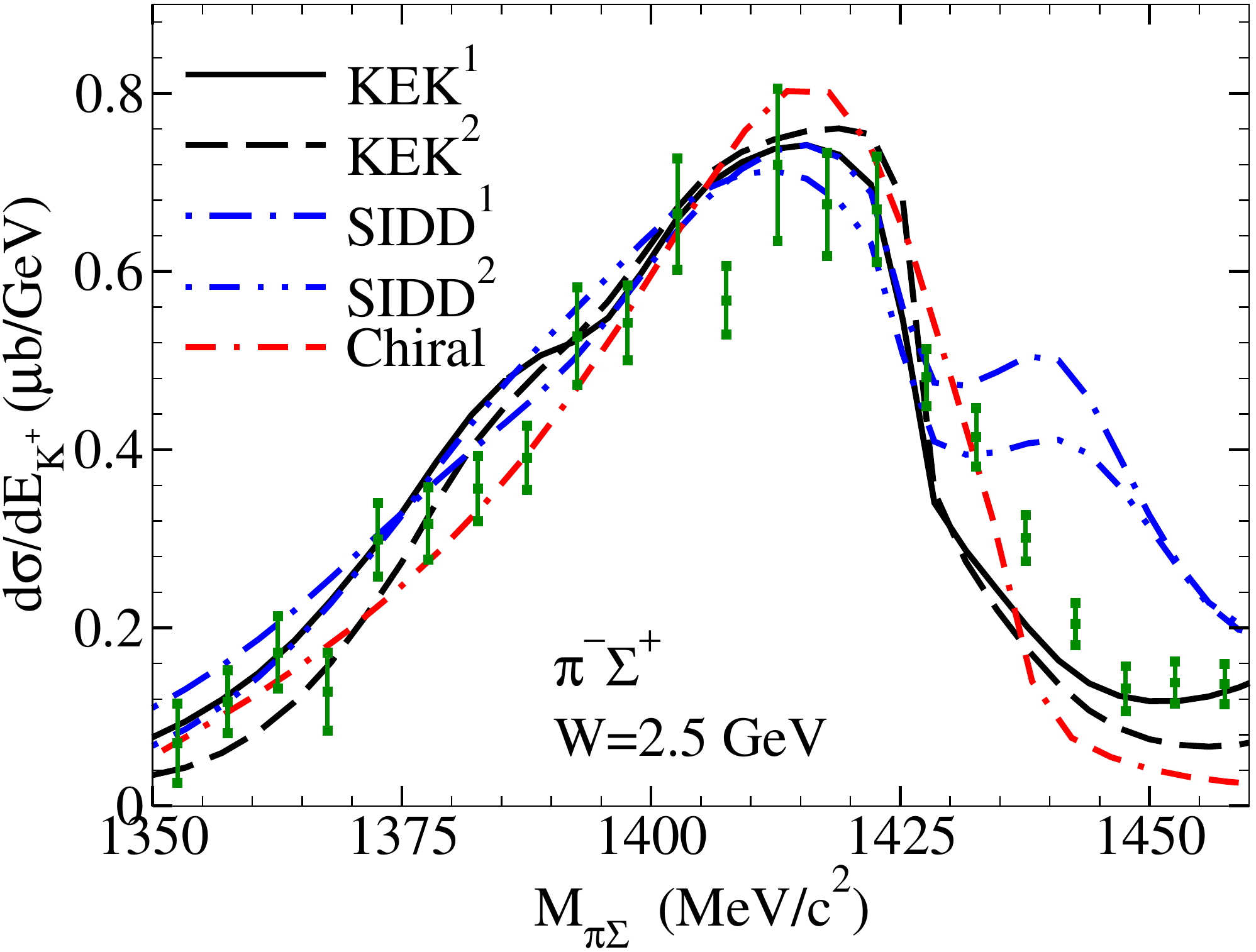}
\hspace{0.2cm}
\includegraphics[scale=0.28]{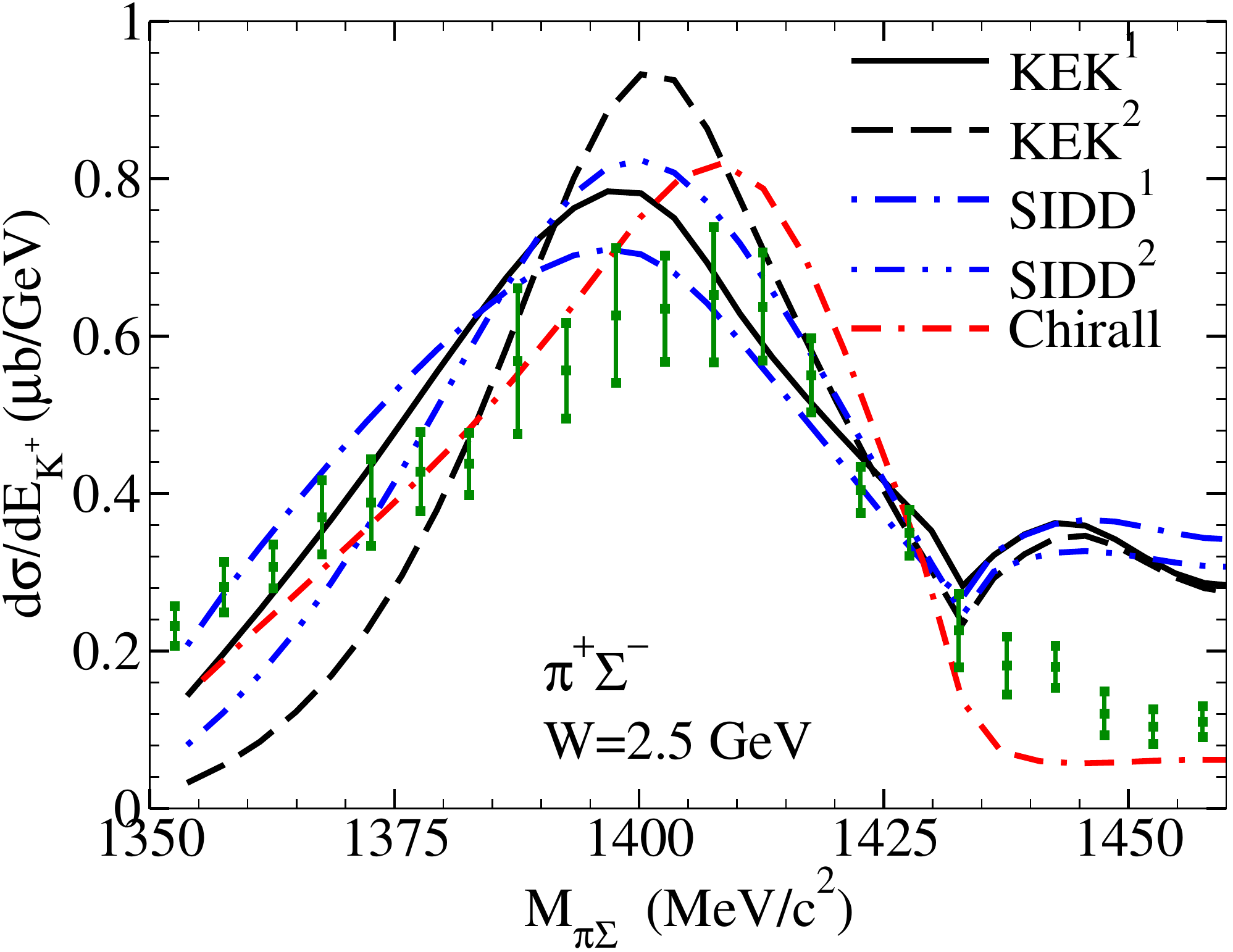} \\
\caption{(Color online) The $\pi\Sigma$ mass spectra for $\gamma{p}\rightarrow(\pi\Sigma)^{0}+K^{+}$ 
reaction. The descriptions are same as in Fig.~\ref{fig.2}, but in the present calculations, 
the full coupled-channel Faddeev AGS equations for $K\bar{K}N-K\pi\Sigma-\pi\pi{N}-\pi\eta{N}$ 
system are solved and all possible initial states of the $K\bar{K}N$ system are included.}
\label{fig.4}
\end{figure*}

\subsection{Dependence of the $\pi\Sigma$ mass spectra on initial channel}
\label{dir3}
The extracted results in Refs.~\cite{jid4,d2,d3} suggest that the $K\bar{K}N$ state can be understood 
by the structure of simultaneous coexistence of $\Lambda(1405)+K$ and $a_{0}(980)+N$ clusters and the 
$\bar{K}$ meson is shared by both $\Lambda(1405)$ and $a_{0}(980)$ at the same time. In Figs.~(\ref{fig.2}) 
and (\ref{fig.3}), it was supposed that the initial state of the $K\bar{K}N$ system is $(\bar{K}N)_{I=0}+K$ 
and the $\pi\Sigma$ mass spectrum was calculated for $(\bar{K}N)_{I=0}+K$ reaction. Now, in present 
subsection,we take into account all possible channels of $K\bar{K}N$ system, namely, $(\bar{K}N)_{I=0,1}+K$, 
$(K\bar{K})^{J=0,1}_{I=0,1}+N$ and $(KN)_{I=0,1}+\bar{K}$. Therefore, the initial state is a mixture of 
the mentioned partitions. The Faddeev equations in (\ref{eq.9}) are related to the initial state 
$K+(\bar{K}N)_{I=0}$. Therefore, by changing the initial state, the Faddeev equations must be changed. 
The Faddeev equations in $\alpha=1$ channel are given by 

\begin{widetext}
\begin{equation}
\begin{split}
& \mathcal{K}^{1,1}_{K,x;I_{{K}},I_{x}J}\,\, =\,\,  (1-\delta_{Kx})\mathcal{M}^{1}_{K,x;I_{K},I_{x}J}
\mathcal{M}^{1}_{K,\bar{K};I_{K},I_{\bar{K}}}
\tau^{11}_{\bar{K};I_{\bar{K}}}\mathcal{K}^{1,1}_{\bar{K},x;I_{\bar{K}},I_{x}J}
+
\mathcal{M}^{1}_{K,N;I_{K},I_{N}}\tau^{11}_{N;I_{N}}\mathcal{K}^{1,1}_{N,x;I_{N},I_{x}J} \\
& \hspace{1.7cm} +\mathcal{M}^{1}_{K,N;I_{K},I_{N}}\tau^{13}_{N;I_{N}} 
\mathcal{K}^{3,1}_{N,x;I_{N},I_{x}J} 
+ \mathcal{M}^{1}_{K,N;I_{K},I_{N}}\tau^{14}_{N;I_{N}}\mathcal{K}^{4,1}_{N,x;I_{N},I_{x}J}\\
& \mathcal{K}^{1,1}_{\bar{K},x;I_{\bar{K}},I_{x}J} \hspace{0.cm}\,\, =\,\, 
(1-\delta_{\bar{K}x})\mathcal{M}^{1}_{\bar{K},x;I_{\bar{K}},I_{x}J}
+
\mathcal{M}^{1}_{\bar{K},K;I_{\bar{K}},I_{K}}\tau^{11}_{K;I_{K}}\mathcal{K}^{1,1}_{K,x;I_{K},I_{x}J}
+
\mathcal{M}^{1}_{\bar{K},K;I_{\bar{K}},I_{K}}\tau^{12}_{K;I_{K}}\mathcal{K}^{2,1}_{K,x;I_{K},I_{x}J} \\
& \hspace{1.7cm}+\mathcal{M}^{1}_{\bar{K},N;I_{\bar{K}},I_{N}}\tau^{11}_{N;I_{N}} 
\mathcal{K}^{1,1}_{N,x;I_{N},I_{x}J} 
+ \mathcal{M}^{1}_{\bar{K},N;I_{\bar{K}},I_{N}}\tau^{14}_{N;I_{N}}\mathcal{K}^{4,1}_{N,x;I_{N},I_{x}J} \\
& \mathcal{K}^{1,1}_{N,x;I_{N},I_{x}J}\,\, =\,\, (1-\delta_{Nx})\mathcal{M}^{1}_{N,x;I_{N},I_{x}J}
+ 
\mathcal{M}^{1}_{N,K;I_{N},I_{K}} \tau^{11}_{K;I_{K}} \mathcal{K}^{1,1}_{K,x;I_{K},I_{x}J} 
+ 
\mathcal{M}^{1}_{N,\bar{K};I_{N},I_{\bar{K}}} \tau^{11}_{\bar{K};I_{\bar{K}}} 
\mathcal{K}^{1,1}_{\bar{K},x;I_{\bar{K}},I_{x}J} \\
& \hspace{1.7cm}+\mathcal{M}^{1}_{N,K;I_{N},I_{K}} \tau^{11}_{K;I_{K}} 
\mathcal{K}^{1,1}_{K,x;I_{K},I_{x}J}.
\end{split}
\label{eq.12}
\end{equation}
\end{widetext}

depending on initial state, the parameters $x$ is the spectator particle in initial state and 
the quantum numbers $I_{x}$ and $J$ are isospin and spin of the interacting pair which are given by
\begin{equation}
\begin{split}
\centering
& x=K:\hspace{0.5cm} K+(\bar{K}N)_{I_{x}J} \hspace{0.5cm} J=0;\,I_{x}=0,1 \\
& x=\bar{K}:\hspace{0.5cm} \bar{K}+(KN)_{I_{x}J} \hspace{0.5cm} J=0;\,I_{x}=0,1 \\
& x=N:\hspace{0.5cm} N+(K\bar{K})_{I_{x}J} \hspace{0.5cm} J=0,1;\,I_{x}=0,1,
\end{split}
\label{eq.13}
\end{equation}
and finally, in $\alpha=2,3,4$ channels, just the indexes of the Faddeev amplitudes must be changed 
\begin{equation}
\mathcal{K}^{\alpha,1}_{i,K;I_{K},0}\rightarrow \mathcal{K}^{\alpha,1}_{i,x;I_{K},I_{x}J}
\label{eq.14}
\end{equation}

Using Eqs.~(\ref{eq.12}), (\ref{eq.13}) and (\ref{eq.14}), we define the scattering amplitude of 
$K\bar{K}N\rightarrow{K}+(\pi\Sigma)^{0}$ reaction as follows,
\begin{equation}
\begin{split}
& T_{(\pi\Sigma){K}\leftarrow(K\bar{K}N)} (\vec{k}_K,\vec{p}_K,P_{x};W) \\
& =\sum_{x,I_{x},J}\{ \sum_{I_{K}}g^{2}_{K;I_{K}}(\vec{k}_K )\tau^{21}_{K;I_{K}}(W-E_K(\vec{p}_K)) \\
&\hspace{4cm}\times\mathcal{K}^{1,1}_{K,x;I_{K},I_{x}J}(p_{K},P_{x};W) \\
& +\sum_{I_{K}}g^{2}_{K;I_{K}}(\vec{k}_K )\tau^{22}_{K;I_{K}}(W-E_{K}(\vec{p}_{K})) \\
&\hspace{4cm}\times\mathcal{K}^{2,1}_{K,x;I_{K},I_{x}J}(p_{K},P_{x};W) \\
& +\sum_{I_{\pi}}\sum_{I_{K}}\langle[\pi\otimes\Sigma]_{I_{K}}\otimes{K}
\mid\pi\otimes[\Sigma\otimes{K}]_{I_{\pi}}
\rangle{g}^{2}_{\pi;I_{\pi}}(\vec{k}_{\pi}) \\
& \hspace{1.5cm}\times \tau^{22}_{\pi;I_{\pi}}(W-E_{\pi}(\vec{p}_{\pi})) 
\mathcal{K}^{21}_{\pi,x;I_{\pi},I_{x}J}(p_{\pi},P_{x};W) \\
& +\sum_{I_{\Sigma}}\sum_{I_{K}}\langle[\pi\otimes\Sigma]_{I_{K}}\otimes{K}
\mid\Sigma\otimes[\pi\otimes{K}]_{I_{\Sigma}}
\rangle{g}^{2}_{\Sigma;I_{\Sigma}}(\vec{k}_{\Sigma}) \\
& \hspace{1.5cm}\times \tau^{22}_{\Sigma;I_{\Sigma}}(W-E_{\Sigma}(\vec{p}_{\Sigma})) 
\mathcal{K}^{21}_{\Sigma,x;I_{\Sigma},I_{x}J}(p_{\Sigma},P_{x};W)\}.
\end{split}
\label{eq.15}
\end{equation}

In Fig.~(\ref{fig.4}), the $\pi\Sigma$ mass spectrum was calculated for different potentials of 
$\bar{K}N-\pi\Sigma$ interaction including all initial channels of $K\bar{K}N$ system. The extracted 
$\chi^{2}$ values are also presented in Table~\ref{ta.2} (the columns that shown by (C) symbol). 
Comparing the results in (C) columns for each model of interaction and scattering energy, one can 
see that the full coupled calculations including all initial channels can better reproduce the 
experimental results. Looking at Table~\ref{ta.2} one can clearly see that for $\mathrm{SIDD}^{1}$ 
potential, the obtained values of the $\chi^{2}$ are considerably smaller than those by other potentials 
for all particle channels and kaon incident energies. Therefore, such a combined study at two different 
initial energies shows a big potential to discriminate between possible mechanisms of the formation 
of $\Lambda(1405)$ resonance.
\subsection{Dependence of the $\chi^{2}$ on $\bar{K}N$ pole position}
\label{sub.2}
In subsection~\ref{dir2}, it was shown that the $\mathrm{SIDD}^{1}$ potential with one-pole nature of 
$\Lambda(1405)$ resonance can reproduce the experimental results with smaller values of $\chi^{2}$. 
In the present subsection, we studied the dependence of the $\chi^{2}$ parameter to the mass and width of 
$\Lambda(1405)$ resonance. To this goal, we constructed a coupled-channel $\bar{K}N-\pi\Sigma$ potential 
having one-pole nature of $\Lambda(1405)$ as a Feshbach resonance~\cite{fesh1,fesh2,fesh3}. To define 
the parameters of the model, we used the following experimental data:
\begin{itemize}
\item Mass and width of the $\Lambda(1405)$ resonances. Assuming that it is a quasi-bound state 
in the $\bar{K}N$ channel and a resonance in the $\pi\Sigma$ channel.
\item The $K^{-}p$ scattering length. For $K^{-}p$ scattering length, we take the reported values 
in Ref.~\cite{baz}
\begin{equation}
a_{K^{-}p}=-0.65+i0.81\,\mathrm{fm}.
\end{equation}
\item The $\gamma$ branching ratio, which is given by
\begin{equation}
\gamma=\frac{\Gamma(K^{-}p\rightarrow\pi^{+}\Sigma^{-})}{\Gamma(K^{-}p
\rightarrow\pi^{-}\Sigma^{+})}=2.36\pm 0.04.
\end{equation}
\item Elastic $K^{-}p\rightarrow K^{-}p$ total cross section.
\end{itemize}

\begin{figure*}[htb]
\vspace{0.5cm}
\centering
\includegraphics[width=8.6cm]{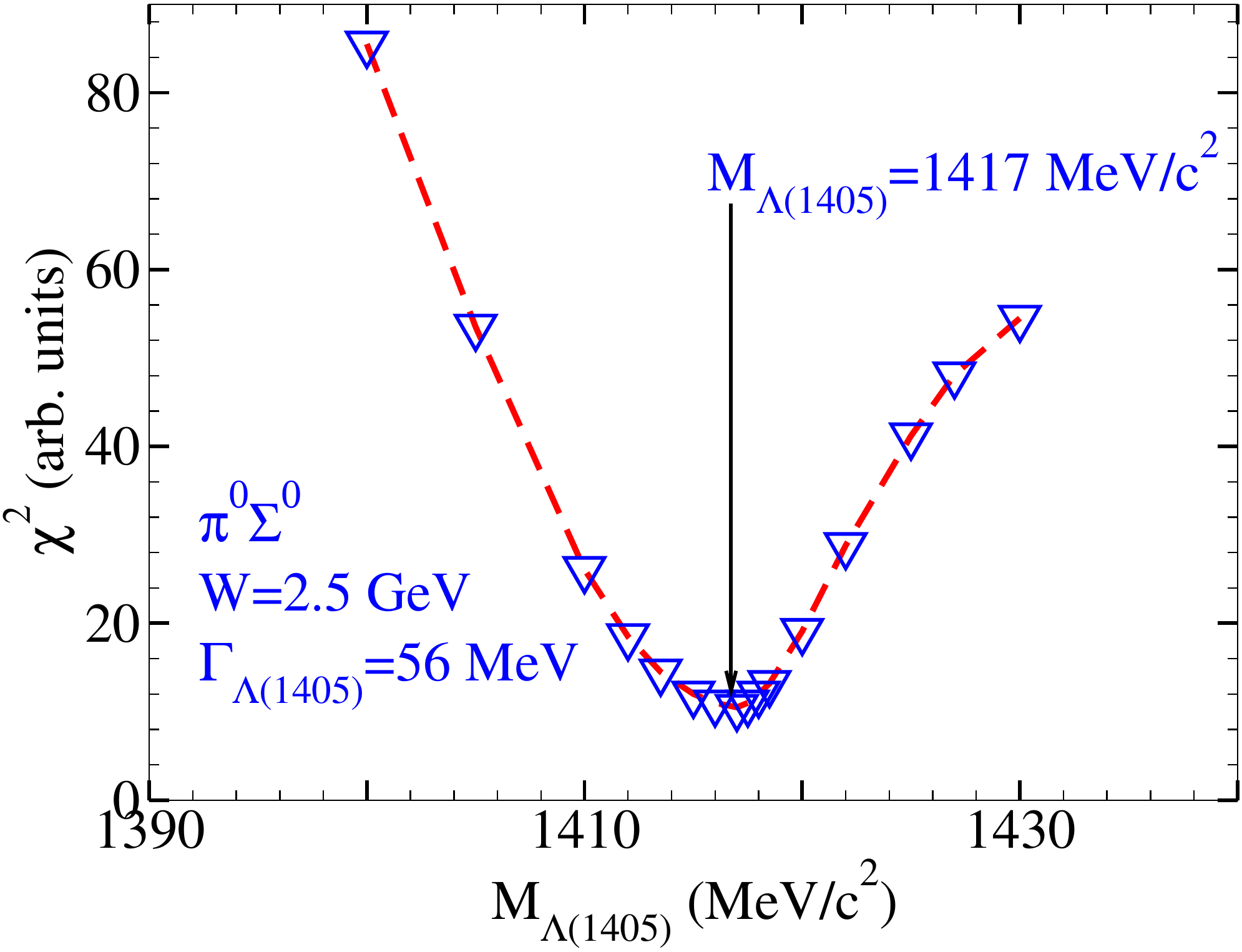}
\hspace{0.5cm}
\includegraphics[width=8.6cm]{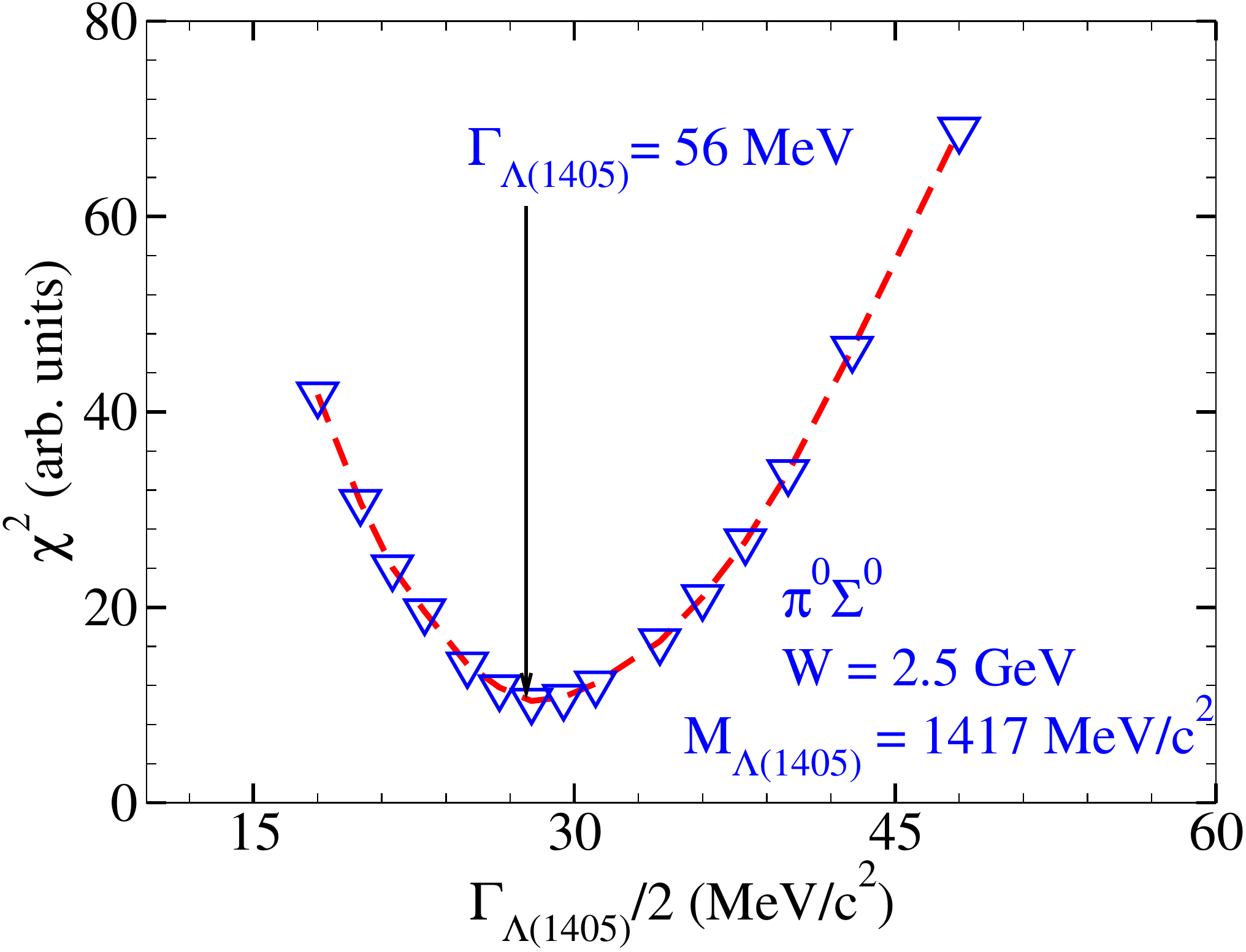}
\caption{(Color online) The dependence of the $\chi^{2}$ value to the real (left panel) and 
imaginary part (right panel) of the $\bar{K}N$ pole position. In the left panel, the imaginary 
part of the pole position is fixed at 28 MeV and the mass of $\Lambda(1405)$ was changed from 
1400 to 1430 MeV/$c^{2}$. In the right panel, the mass is fixed at 1417 MeV/$c^{2}$ and the 
width was changed from 30 to 100 MeV. }
\label{fig.5}
\end{figure*}

We studied the dependence of the $\pi^{0}\Sigma^{0}$ mass spectrum to the $\bar{K}N$ pole position 
at $W=2.5\mathrm{GeV}$. In Fig.~(\ref{fig.5}), the variation of the $\chi^{2}$ values with respect 
to the real (left panel) and imaginary (right panel) part of the $\bar{K}N$ pole position for 
$\pi^{0}\Sigma^{0}$ mass spectra at energy $W=2.5\mathrm{GeV}$ are depicted. In the left panel, the 
imaginary part of the $\bar{K}N$ pole position is fixed to be 28 MeV and in the right panel the real 
part of the pole position is 1417 MeV. As one can see, in the left panel, the mass of 1417 
$\mathrm{MeV/c^{2}}$ gives us the minimum value of the $\chi^{2}$ parameter and in the right panel 
the width of 56 $\mathrm{MeV}$ was extracted for $\Lambda(1405)$ resonance.

In the present calculations, the best fit was extracted just by taking into account the effect of 
$\Lambda(1405)$ resonance but, there are other factors which may also have important effect on the 
obtained results, here. The first one is $\Sigma(1385)$ resonance, to be more precise, one should 
include also the effect of $\Sigma(1385)$ resonance which may change the extracted results. Especially, 
to find a better fit in energy region below 1400 MeV, the inclusion of $\Sigma(1385)$ resonance could 
be important. One can extend these calculations to include the effect of $\Sigma(1385)$ resonance. 

The kinetic energy of the system under consideration in this calculation is between 100$\sim$600 MeV 
in $K\bar{K}N$ channel and in lower-lying channels is more higher. comparing this kinetic energy with 
the averaged mass of pion 139 MeV/$c^{2}$, therefore, the inclusion of relativistic effects will also 
be important and the relativistic corrections can change the extracted results in present work. The 
two-body interactions are driven nonrelativisticaly. Therefore, in the present calculations, the 
nonrelativistic Faddeev equations were solved for $K\bar{K}N$ system. One can also include the 
relativistic corrections by employing the methods presented in Refs.~\cite{re1,re2}.
\section{CONCLUSION}
\label{conc}
In the present calculations, we solved the Faddeev AGS equations for $K\bar{K}N$ system. The 
$\pi\Sigma$ invariant mass was calculated for all combinations of charges, i.e., $\pi^{-}\Sigma^{+}$, 
$\pi^{+}\Sigma^{-}$ and $\pi^{0}\Sigma^{0}$ based on this approach. To study the dependence 
of the $\pi\Sigma$ mass spectrum to the two-body interactions, different types of $\bar{K}N$ 
and $K\bar{K}$ interactions were used. We calculated the $\chi^{2}$ value for all models of 
interaction. It was shown that the mass spectrum resulting from the one-pole version of the 
$\bar{K}N-\pi\Sigma$ potential is in more agreement with the experimental results for 
$\gamma{p}\rightarrow {K}^{+}(\pi\Sigma)^{0}$ reaction. However, the extracted results do not 
confidently say which model of interaction can describe the structure of $\Lambda(1405)$ 
resonance, exactly. 

In the present study, the effect of different factors on the final $\pi\Sigma$ mass spectrum was 
studied. By solving the full coupled-channel Faddeev equations for 
$K\bar{K}N-K\pi\Sigma-\pi\pi{N}-\pi\eta{N}$ system, the dependence of the mass spectrum on 
$t_{\pi\Sigma\leftarrow\pi\Sigma}$ was studied. The effect of different initial channels on 
mass spectrum and $\chi^{2}$ values was investigated and it was shown that in exact study of 
CLAS data, one should consider the effect of these factors. We also used the $\Lambda(1405)$ 
pole position and width as fitting parameters and as shown in Fig.~(\ref{fig.5}), a minimum 
is observed at around $M_{\Lambda(1405)}=1417$ $\mathrm{MeV}/c^{2}$ and $\Gamma_{\Lambda(1405)}=56$ 
MeV in the $\chi^{2}$ distribution. 

\end{document}